\documentclass{aa}
\usepackage{natbib}
\usepackage{xspace}
\usepackage{amssymb}
\usepackage{amsmath}
\usepackage{graphicx}

\def\revised#1{{#1}}

\def\aap{A\&A}

\def\mnras{MNRAS}

\def\apj{ApJ}

\def\apjl{ApJL}

\def\gram{\hbox{g}}
\def\cm{\hbox{cm}}

\def\yr{\hbox{yr}}
\def\AU{\hbox{AU}}

\def\comma{\,,}
\def\fullstop{\,.}
\def\tfric{\ensuremath{t_\mathrm{fric}}\xspace}
\def\teddy{\ensuremath{t_\mathrm{edd}}\xspace}
\def\tset{\ensuremath{t_\mathrm{sett}}\xspace}
\def\tdif{\ensuremath{t_\mathrm{stir}}\xspace}
\def\tkep{\ensuremath{t_\mathrm{kep}}\xspace}
\def\taccr{\ensuremath{t_\mathrm{accr}}\xspace}
\def\veddy{\ensuremath{v_\mathrm{edd}}\xspace}
\def\leddy{\ensuremath{l_\mathrm{edd}}\xspace}
\def\ffric{\ensuremath{F_\mathrm{fric}}\xspace}
\def\vset{\ensuremath{v_\mathrm{sett}}\xspace}
\def\fgrav{\ensuremath{F_\mathrm{grav}}\xspace}
\def\zstir{\ensuremath{z_{\mathrm{sett}}}\xspace}
\def\zdepl{\ensuremath{z_{\mathrm{depl}}}\xspace}
\def\rturn{\ensuremath{R_{\mathrm{turn}}}\xspace}
\def\fact{\ensuremath{\xi}\xspace}
\def\cs{\ensuremath{c_\mathrm{s}}\xspace}
\def\sigcoll{\ensuremath{\sigma}\xspace}
\def\Tgas{\ensuremath{T_\mathrm{gas}}\xspace}
\def\rhogas{\ensuremath{\rho}\xspace}
\def\rhodust{\ensuremath{\rho_\mathrm{d}}\xspace}
\def\mugas{\ensuremath{\mu_\mathrm{gas}}\xspace}
\def\mp{\ensuremath{m_\mathrm{p}}\xspace}
\def\Hp{\ensuremath{H_\mathrm{p}}\xspace}
\def\Hs{\ensuremath{H_\mathrm{s}}\xspace}
\def\Omegak{\ensuremath{\Omega_\mathrm{K}}\xspace}
\def\Mstar{\ensuremath{M_\star}\xspace}
\def\Sc{\ensuremath{\mathrm{Sc}}\xspace}
\def\St{\ensuremath{\mathrm{St}}\xspace}
\def\Pt{\ensuremath{\mathrm{Pt}}\xspace}
\begin{document}
\title{The effect of dust settling on the appearance of protoplanetary disks}
\titlerunning{Dust settling in protoplanetary disks}
\authorrunning{Dullemond \& Dominik}
\author{C.P.~Dullemond \& C.~Dominik}
\institute{Max Planck Institut f\"ur Astrophysik, P.O.~Box 1317, D--85741 
Garching, Germany; e--mail: dullemon@mpa-garching.mpg.de\\
Sterrenkundig Instituut `Anton Pannekoek', Kruislaan 403,
  NL-1098 SJ Amsterdam, The Netherlands; e--mail: dominik@science.uva.nl}
\date{DRAFT, \today}

\abstract{We analyze how the process of dust settling affects the spectral
  energy distribution and optical appearance of protoplanetary disks.  Using
  simple analytic estimates on the one hand, and detailed 1+1-D models on
  the other hand, we show that, while the time scale for settling down to
  the equator may exceed the life time of the disk, it takes much less time
  for even small grains of 0.1 $\mu$m to settle down to a few pressure scale
  heights. This is often well below the original location of the disk's
  photosphere, and the disk therefore becomes effectively 'flatter'.  If
  turbulent stirring is included, a steady state solution can be found,
  which is typically reached after a few $\times$ 10$^5$ years.  In this
  state, the downward settling motion of the dust is balanced by vertical
  stirring. Dependent on the strength of the turbulence, the shape of the
  disk in such a steady state can be either fully flaring, or flaring only
  up to a certain radius and self-shadowed beyond that radius. These
  geometries are similar to the geometries that were found for disks around
  Herbig Ae/Be stars in our previous papers (Dullemond
  \citeyear{dullemond:2002}; Dullemond \& Dominik A\&A in press, henceforth
  DD04). In those papers, however, the reason for a disk to turn
  self-shadowed was by loss of optical depth through dust grain growth. Here
  we show that dust settling can achieve a similar effect without loss of
  vertical optical depth, although the self-shadowing in this case only
  affects the outer regions of the disk, while in DD04 the entire disk
  outside of the puffed-up inner rim was shadowed.  In reality it is likely
  that both grain growth and grain settling act simultaneously. The spectral
  energy distributions of such self-shadowed --- or partly self-shadowed ---
  disks have a relatively weak far-infrared excess (in comparison to flaring
  disks). We show here that, when dust settling is the cause of
  self-shadowing, these self-shadowed regions of the disk are also very weak
  in resolved images of scattered light. A reduction in the brightness was
  already predicted in DD04, but we find that dust settling is far more
  efficient than grain growth at dimming the scattered light image of the
  disk. Settling is also efficient in steepening the \revised{spectral
  energy distribution} at \revised{mid- to} far-infrared wavelengths. From
  the calculations with compact grains it follows that, after about 10$^6$
  years, most disks should be self-shadowed. The fact that some older disks
  are still observed with the characteristics of flaring disks therefore
  seems somewhat inconsistent with the time scales predicted by the settling
  model based on compact grains. This suggests that perhaps even the small
  grains ($\lesssim 0.1\mu$m) have a porous or fractal structure, slowing
  down the settling.  \revised{Or it could mean that the different
  geometries of observed disks is merely a reflection of the turbulent state
  of these disks.}}

\maketitle


\section{Introduction}

The dusty circumstellar disks surrounding most T Tauri stars and Herbig
Ae/Be stars are believed to be the birthplace of planetary systems.  These
systems are therefore the object of intense study, both observationally and
theoretically.  Much of what we know of these systems has been derived
indirectly from their spectral energy distributions and infrared spectra.
In the last few years the amount of observational information on spatial
structure of these objects has increased dramatically.  The outer disk
regions (typically outside of 50 AU) can be probed by direct imaging, both
at optical/near-infrared wavelengths with HST/STIS (e.g.~Augereau et
al.~\citeyear{augereau:2001}; Grady et al.~\citeyear{gradywoodbruh:1999}),
and in the sub-mm regime (Mannings \& Sargent \citeyear{mannsarg:1997}), and
with near- and \revised{mid-infrared} interferometry the inner disk
structures can be probed (e.g.~Eisner et al.~\citeyear{eislanake:2003};
Millan-Gabet et al.~\citeyear{millanschl:2001}; Leinert et al.~A\&A subm.).

Models of protoplanetary disks are increasingly successful at accounting for
much of the observed properties.  For instance, they can explain why their
\revised{spectral energy distributions (SEDs)} are generally rather flat in
$\nu F_\nu$ (Kenyon \& Hartmann \citeyear{kenyonhart:1987}), why dust
features are almost always observed in emission (Calvet et
al.~\citeyear{calvetpatino:1991}; Chiang \& Goldreich
\citeyear{chianggold:1997}), what the meaning is of the \revised{near-infrared} bump
in the SEDs of Herbig Ae stars (Natta et al.~\citeyear{nattaprusti:2001};
Dullemond, Dominik \& Natta \citeyear{duldomnat:2001}), and why some sources
have strong and some have weak far-IR excess (Dullemond
\citeyear{dullemond:2002}; Dullemond \& Dominik A\&A in press, henceforth
DD04).

For the overall disk properties, it is often sufficient to work with a disk
model in which dust and gas are well mixed and coupled.  However, there are
indications that dust growth, thermal processing, and dust-gas separation
can play an important role in disks. The different overall shapes of the
SEDs of Herbig AeBe stars (the ``group I'' and ``group II'' spectra
identified by Meeus et al.~\citeyear{meeuswatersbouw:2001}) can be
understood by a reduction in opacity in the outer disk via grain growth
(DD04; see also D'Alessio et al.~\citeyear{dalessiocalvet:2001}) or a direct
reduction of the disk height by settling of small grains (Chiang et
al.~\citeyear{chiangjoung:2001}). The high submm fluxes observed in disks
with relatively low far-IR fluxes can be explained by a component of large
grains in the disk interior which do not contribute to the opacity high
above the midplane but contribute only as a cold component to the submm
emission (Natta et al.~\citeyear{nattaprusti:2001}; DD04). Both grain growth
and grain settling is required to understand the presence of these grains.
Moreover, the infrared spectra taken from the ground and from space have
revealed structure in the infrared dust emission features that are
indicative of dust grain growth and thermal processing (Bouwman
\citeyear{bouwmanmeeus:2001}; van Boekel \citeyear{vanboekelwaters:2003};
Honda et al.~\citeyear{hondakataza:2003}; Meeus et
al.~\citeyear{meeussterz:2003}; Przygodda et
al.~\citeyear{przygoddaboek:2003}).

Also from a theoretical point of view, grain processing and dust-grain
separation are expected to occur. In fact, a central role in our current
understanding of planet formation is played by the formation of a \emph{dust
subdisk} within the disk \citep{safronov-book,dubmorster:1995}.  Dust grains
tend to settle towards the midplane, unless this process is countered by
strong vertical stirring in the disk.  The formation of such a subdisk with
enhanced dust density greatly influences the timescales in which dust grains
may grow to planetesimals and eventually planets.  For the detailed
structure of the dust subdisk, parameters like the remaining accretion rate,
local turbulence, radial mixing processes, etc.~play and important role
\citep{dubmorster:1995,2003Icar..166..385C}.  Unfortunately, the processes
that generate turbulence in protoplanetary disks are not well understood,
and mixing/stirring is generally treated in a simple parameterized manner.

One way to approach this problem is to search for observational signs of
dust settling in disks.  Observations of disks, at least in the optical and
infrared, mostly probe the upper disk layers.  If settling is active in
disks, the upper layers of a disk will be influenced and may show observable
changes.  While there are many studies into the effects of settling onto the
dust distribution relevant for planet formation (see e.g.~Takeuchi \& Lin
\citeyear{takeuchilin:2002}, \citeyear{takeuchilin:2003}), there are
curiously few studies which discuss the observational consequences of dust
settling in a quantitative way.  A study of this kind was presented by
Miyake and Nakagawa (\citeyear{miyakenaka:1995}, hereafter MN95). They
studied the effects of grain settling and accretion luminosity on the SED of
T Tauri disks in order to explain the infrared and IRAS colors of these
stars.  However, their study is limited in several ways.  MN95 did not
consider turbulent stirring at all. The basic assumption of their model was
that turbulence has completely ceased and dust settling can proceed
undisturbed. The dust was assumed to be confined within one pressure scale
height, and the time scales for changing the surface height in the disk were
computed using the midplane density.  These time scales turn out to be
between 10$^5$ and 10$^{7}$ years, and therefore comparable to the lifetime
of protoplanetary disks.  However, the timescales for settling are much
shorter in the higher layers of the disk. Moreover, we will show that
turbulence prevents the dust from settling down below a certain height,
often much larger than the pressure scale height, and that turbulence will
therefore determine the final shape of the photospheric surface of the disk.

In this paper we investigate the effects of dust settling in the presence of
turbulent stirring on the observable quantities of protoplanetary disks, in
particular their SEDs and their images in scattered light. To do this we
start with a well-mixed disk model for a T Tauri star and compute how the
process of dust settling proceeds. In all stages of this process we compute
observable quantities and see how they change.  In doing this we hope to
find typical observable signatures of dust settling which can be used to
analyze protoplanetary disks.

The paper is organized as follows: In section~\ref{sec-estimates} we
use simple analytical estimates to derive the timescales for settling
and stirring.  In section~\ref{sec-vert-models} we solve the equations
for settling and stirring in a one-dimensional slab, a vertical cut
through the disk and show how quickly the upper layers of the disk are
depleted and to what depth.  In section~\ref{sec:1+1-d-disk} we run a
similar calculation at all locations in the disk and compute the
changes settling causes to both the SED and images in scattered
light.

\section{Settling, vertical stirring and accretion}\label{sec-estimates}
The equations for settling, vertical stirring and accretion have been
extensively discussed in the literature.  But, for the sake of
clarity, and to define our notation, we include a brief discussion of
the entire set of equations used for the current study.

Dust grains in a protoplanetary disk experiences a gravitational force
towards the star.  If it were not for the gas in the disk, the grains would
move on inclined orbits.  In a co-rotating frame this means they would
oscillate about the midplane with the Keplerian frequency.  However, small
grains have a large surface-to-mass ratio, and therefore experience strong
drag forces with the gas. Instead of oscillating about the midplane, these
grains slowly settle toward the midplane on a time scale of a few hundred
thousand years. This time scale depends on the surface density of the gas
and on the surface-to-mass ratio of the grain.  Large compact grains have a
much smaller surface-to-mass ratio and therefore tend to settle on a much
shorter time scale. Settling can, however, only take place when turbulence
in the disk does not stir up the grains above the midplane again. Usually
this happens up to a certain height $\zstir$ above the midplane while above
this height the settling can proceed without problems. This means that after
a certain settling time, the disk above $\zstir$ is more or less devoid of
dust grains of this particular size, while below it the vertical stirring
has more or less equalized the abundance of dust grains of this size.

In this section we derive the equations for these processes, and estimate
the height $\zstir$ and the time scale that it takes for the dust to settle
down to this height.

\subsection{Settling}
The interaction between a dust grain and the surrounding gas can be
described by the \emph{friction time} which is the time scale in which a
dust grain responds to the motion of the gas.
\begin{equation}
\label{eq:1}
\tfric=\frac{mv}{|\ffric|}
\end{equation}
where $m$ is the mass of the dust particle and $v$ the relative velocity
between gas and dust.  The friction force \ffric is a function of the drift
velocity $v$, the grain properties and the gas density and temperature.  In
general, different regimes must be distinguished depending on the
hydrodynamic regime appropriate for a given situation.  However, for grains
smaller than typically 1\,cm, the Epstein regime is applicable throughout a
protoplanetary disks (Cuzzi et al.~\citeyear{cuzzidobrchamp:1993}), so we
restrict ourselves for the current paper to this regime.  Then, under the
generally valid assumption of subsonic drift velocities, the friction force
is given by
\begin{equation}
\label{eq:2}
\ffric = - \frac{4}{3}\rhogas\sigcoll v \cs
\end{equation}
where \rhogas is the gas density, \sigcoll is the collisional cross
section of the dust grain and \cs is the isothermal sound velocity
\begin{equation}
\label{eq:3}
\cs=\sqrt{\frac{k\Tgas}{\mugas\mp}} \fullstop
\end{equation}
\Tgas is the gas temperature, $k$ the Boltzman constant, \mugas the
mean molecular weight in the gas and \mp the proton mass.  In the
Epstein regime, the friction time can be written as
\begin{equation}
\label{eq:4}
\tfric=\frac{3}{4}\frac{m}{\sigcoll}\frac{1}{\rhogas\cs} \fullstop
\end{equation}
The friction time depends upon the dust grain properties through the
mass-to-crosssection ratio $m/\sigcoll$.  For compact spherical grains with
grain radius $a$ and specific density \rhodust, $m/\sigcoll=4\rhodust a/3$,
and the friction time turns into the familiar form $\tfric=\rhodust
a/\rhogas\cs$.  However, through much of this paper we will continue to
write $m/\sigcoll$ since this expression is much more general and allows for
a direct generalization towards 2-dimensional grains (like \revised{polycyclic 
aromatic hydrocarbons (PAHs)}) and porous or fractal grains.

We now consider a circumstellar disk around a star with mass \Mstar.
We assume the disk to be isothermal in the vertical direction, \revised{and
we make the thin-disk-approximation: $z\ll R$}.  At a
distance $R$ from the star, the vertical density distribution is then
given by 
\begin{equation}
\label{eq:5}
\rhogas(z) = \frac{\Sigma}{\sqrt{2\pi}\Hp}
 \exp\left(\frac{-z^2}{2\Hp^2}\right)
\end{equation}
where $\Sigma$ is the surface density in the disk, \Hp the pressure
scale height and $z$ the height above the disk.  The pressure scale
height is given by
\begin{equation}
\label{eq:6}
\Hp=\frac{\cs}{\Omegak}
\end{equation}
where $\Omegak=\sqrt{G\Mstar/R^3}$ denotes the local Kepler
frequency. A dust grain located at $(R,z)$ is subject to a vertical
gravitational force
\begin{equation}
\label{eq:7}
\fgrav = - m \Omegak^2z
\end{equation}
and, in the absence of turbulent stirring, settles towards the midplane
at a settling speed
\begin{equation}
\label{eq:8}
\vset = - z \Omegak^2\tfric=
- \frac{3\Omegak^2z}{4\rhogas\cs}\frac{m}{\sigcoll} \comma
\end{equation}
\revised{which follows from $\fgrav = \ffric$.}

The settling speed decreases as one comes closer to the equator,
because the gas density increases and the vertical component of the
gravity decreases.  The time for a grain to settle to a certain height
$z$ is therefore dominated by the local settling speed at height $z$
\begin{equation}
\label{eq:9}
\tset = \frac{z}{\vset} =\frac{4}{3}\frac{\sigcoll}{m}\frac{\rhogas\cs}{\Omegak^2}
\fullstop
\end{equation}
Using the vertical density distribution in Eq.\eqref{eq:5}, we find
\begin{equation}
\label{eq:10}
\tset = \frac{4}{3 \sqrt{2\pi}}\frac{\sigma}{m}\frac{\Sigma}{\Omegak}
\exp \left( - \frac{z^2}{2\Hp^2} \right) \fullstop
\end{equation}
The time for a grain to settle towards the midplane of a disk is therefore
approximately given by $\tset=(4/3\sqrt{2\pi})(\sigcoll/m)(\Sigma/\Omegak)$.
However, it is important to know, that in the higher layers of the disk, the
timescales are very different.  Typical flaring disk models intercept
stellar radiation at a height $z\sim 4\Hp$.  The local settling time scales
in this region differ from the time scales close to the midplane by a factor
$e^{-8}\approx 1/3000$.  Therefore, dust settling will deplete the upper
layers of a disk surface much faster than the actual formation time of the
dust subdisk close to the midplane (see also Chiang \& Goldreich
\citeyear{chianggold:1997}; Takeuchi \& Lin \citeyear{takeuchilin:2002}).

\subsection{Vertical stirring}
Vertical stirring occurs when the disk is turbulent. The turbulent eddies of
the gas transport the dust grains up and down in a random manner, thus
constituting a diffusion process for the dust. A detailed formulation of the
theory behind this turbulent diffusion of dust grains is given by
e.g.~V\"olk et al.~(\citeyear{voelkmorroejon:1980}); Weidenschilling
(\citeyear{weid1997}); Cuzzi, Dobrovolskis \& Champney
(\citeyear{cuzzidobrchamp:1993}); Dubrulle, Morfill \& Sterzik
(\citeyear{dubmorster:1995}), and most recently by Schr\"apler \& Henning
(A\&A subm.). 

The diffusion coefficient for gas molecules $D_0$ is defined using the
standard $\alpha$ prescription:
\begin{equation}
D_0 = \alpha \cs\Hp
\fullstop
\end{equation}
But dust particles may not always couple perfectly to the gas. The actual
diffusion coefficient for dust particles of a certain size is:
\begin{equation}
D = \frac{D_0}{\Sc}
\comma
\end{equation}
where $\Sc$ is the Schmidt number. For perfectly coupled particles
(i.e.~infinitely small grains) the Schmidt number is unity. The decoupling
from the turbulence is described by a Schmidt number that is larger than
unity. Various expressions for the Schmidt number have appeared in the
literature. We adopt:
\begin{equation}
\Sc = (1+\St)
\comma
\end{equation}
where $\St$ is the Stokes number.
The Stokes number $\St$ is defined as the ratio of the friction time
$\tfric$ and the eddy turn-over time $\teddy$. The latter depends not only
on the turbulence parameter $\alpha$, but also on the typical velocity of
the largest eddies. This eddy velocity may depend on the type of turbulence
(e.g.~shear turbulence, convective turbulence, magneto-rotational
turbulence), and it is not clearly agreed in the literature what it should
be. Therefore we parameterize it in the following way:
\begin{equation}
\veddy = \alpha^q \cs
\comma
\end{equation}
where $\veddy$ is the average velocity of the largest eddies and $q$ a
`turbulence parameter' between $0$ and $1$. Schr\"apler \& Henning take
$q=1/2$ in their calculations. The typical largest eddy size $\leddy$ can be
found by the definition that $D_0 \equiv \leddy\veddy$. One obtains
$\leddy=\alpha^{1-q}\Hp$. This leads us to an expression for the eddy
turn-over time $\teddy\equiv \leddy/\veddy =\alpha^{1-2q}/\Omegak$. The
Stokes number then becomes:
\begin{equation}
\St = \frac{\tfric}{\teddy} =
\frac{3}{4}\frac{m}{\sigma}\frac{\Omegak}{\rho\cs}\alpha^{2q-1} \fullstop
\end{equation}
In this paper we will take $q=1/2$, following Schr\"apler \& Henning.

The turbulent $\alpha$ parameter used here is proportional to the
$\alpha_{\mathrm{accr}}$ of the accretion process in the disk. Traditionally
this is denoted as:
\begin{equation}
\alpha = \Pt \; \alpha_{\mathrm{accr}}
\end{equation}
where $\Pt$ is the Prandtl number, which is a dimensionless constant of
order unity. In this paper we assume it to be $\Pt=1$ for the vertical
turbulence.

\subsection{Equilibrium between stirring and settling}
High above the midplane the densities of the gas are so low that turbulence
is not able to effectively mix the grains upward. Instead the grains settle
to the midplane unhindered with a velocity given by Eq.~(\ref{eq:8}). At
some height $z$ above the equator the densities become high enough for the
turbulent eddies to start to counteract the settling.  This height can be
estimated by comparing the time scale for settling $\tset$ to the time scale
for diffusion $\tdif=z^2/D$ which obeys
\begin{equation}
\tdif = \frac{\Sc}{\alpha \Omega_K} \frac{z^2}{H_p^2}
\fullstop
\end{equation}
By equating $\tdif=\fact\tset$ (where $\fact$ is a factor described below)
one obtains the following equation in $z$:
\begin{equation}\label{eq-zsett}
\frac{H_p^2}{z^2}\exp\left(-\frac{z^2}{2H_p^2}\right) = 
\frac{3\sqrt{2\pi}}{4}\frac{m}{\sigma}\frac{\Sc(z)}{\fact\alpha\Sigma}
\end{equation}
By taking $\fact=1$ and solving for $z$ (taking into account that $\Sc$ is a
function of $z$), one obtains the height $\zstir/R$ down to which the grains
of a typical size and mass tend to settle. Below this height the turbulent
diffusion manages to sweep up the grains and keep them afloat. The abundance
for grains of this size is almost constant below this height. Above this
height the abundance decreases because dust can settle down to lower
altitude. It should be noted, however, that the transition from fully mixed
to fully depleted is not extremely sharp. As will be shown in Section
\ref{sec-vert-models}, by solving Eq.(\ref{eq-zsett}) with $\fact=100$ one
obtains a good estimate of the `depletion height' $\zdepl$ above which the
turbulence is so ineffective that dust grains of this size have depleted
almost entirely. The transition from well-mixed to dust-free takes place
between $\zstir$ and $\zdepl$. \revised{Incidently, it is interesting to
note that for grains smaller than 100 $\mu$m the Schmidt number $\Sc$ is
almost always close to $1$ at $\zdepl$ and below. The decoupling of the
grains from the turbulence therefore has only marginal effect on the 
solutions.}

\subsection{Steady state analytic estimates}\label{sec-analytic-est}
Based on the above equations one can now make figures of these quantities
for a particular disk model. Consider a T Tauri star of mass
$M_{*}=0.5M_{\odot}$, radius $R_{*}=2.5R_{\odot}$ and effective temperature
$T_{*}=4000$K. Surround it with a disk with a gas+dust surface density
$\Sigma(R)$ given by
\begin{equation}
\Sigma(R) = \Sigma_0 \left(\frac{R}{1\,\mathrm{AU}}\right)^p
\end{equation}
with $p=-1.5$ and $\Sigma_0=400\gram/\cm^2$, with an inner radius of
$R_{\mathrm{in}}=0.1\AU$ and and out radius $R_{\mathrm{out}}=400\AU$.  The
disk mass, for this configuration, is $M_{\mathrm{disk}}=0.011\,M_{\odot}$.
We will assume, for the time being, that the disk is vertically isothermal,
with a temperature $T(R)$ that is determined by the irradiation of the disk
by the central star. \revised{We ignore viscous heating, but we discuss its
effect in Subsection \ref{sec-visc-diss}.} In order to keep things simple
and analytic, we {\em assume} that the flaring angle $\phi$ (see Chiang \&
Goldreich \citeyear{chianggold:1997} for the definition) is a constant, and
has a value of $\phi=0.05$, which is not too far from realistic values. By
equating the irradiation flux $F_{\mathrm{irr}}=\phi L_{*}/(4\pi R^2)$ to
the cooling flux $F_{\mathrm{cool}}=\sigma_s T(R)^4$ one can solve directly
for $T(R)$, obtaining:
\begin{equation}\label{eq-temp-irrad}
T(R) = \phi^{1/4}\sqrt{\frac{R_{*}}{R}}\, T_{*} \fullstop
\end{equation}
We ignored geometric effects of order $z/R$. Given the temperature $T(R)$,
the vertical density structure is then given by Eqs.(\ref{eq:3}, \ref{eq:5},
\ref{eq:6}).  Note that this disk structure is highly simplified, and in
particular the assumption that $\phi$ is a positive constant is often
untrue. But for the purpose of this paper it is sufficient. \revised{
Throughout this paper the dust specific weight is taken to be $3.6$
g/cm$^3$ and the mean molecular mass is taken $\mu=2.3$.}

\begin{figure*}
\centerline{
\includegraphics[width=9cm]{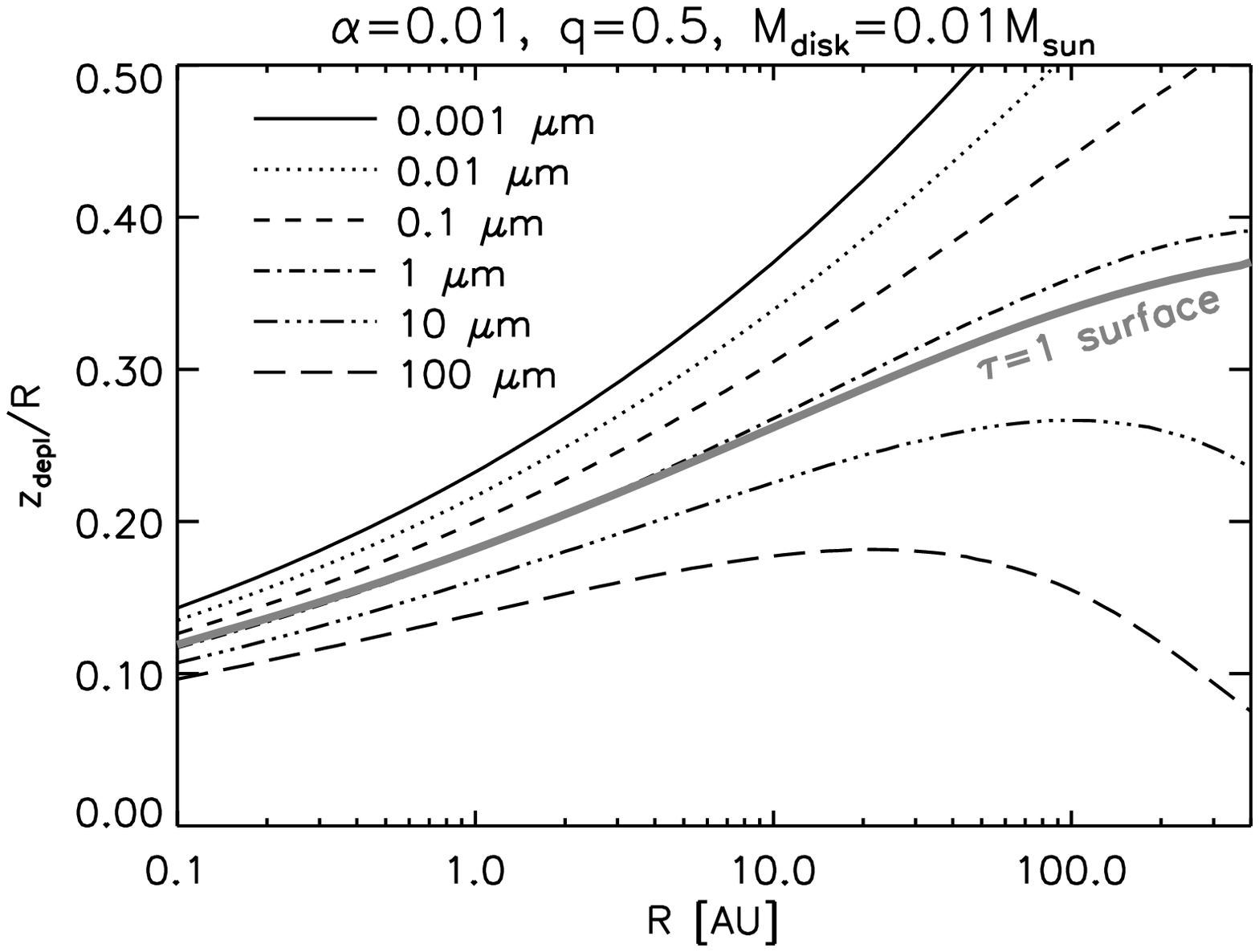}
\includegraphics[width=9cm]{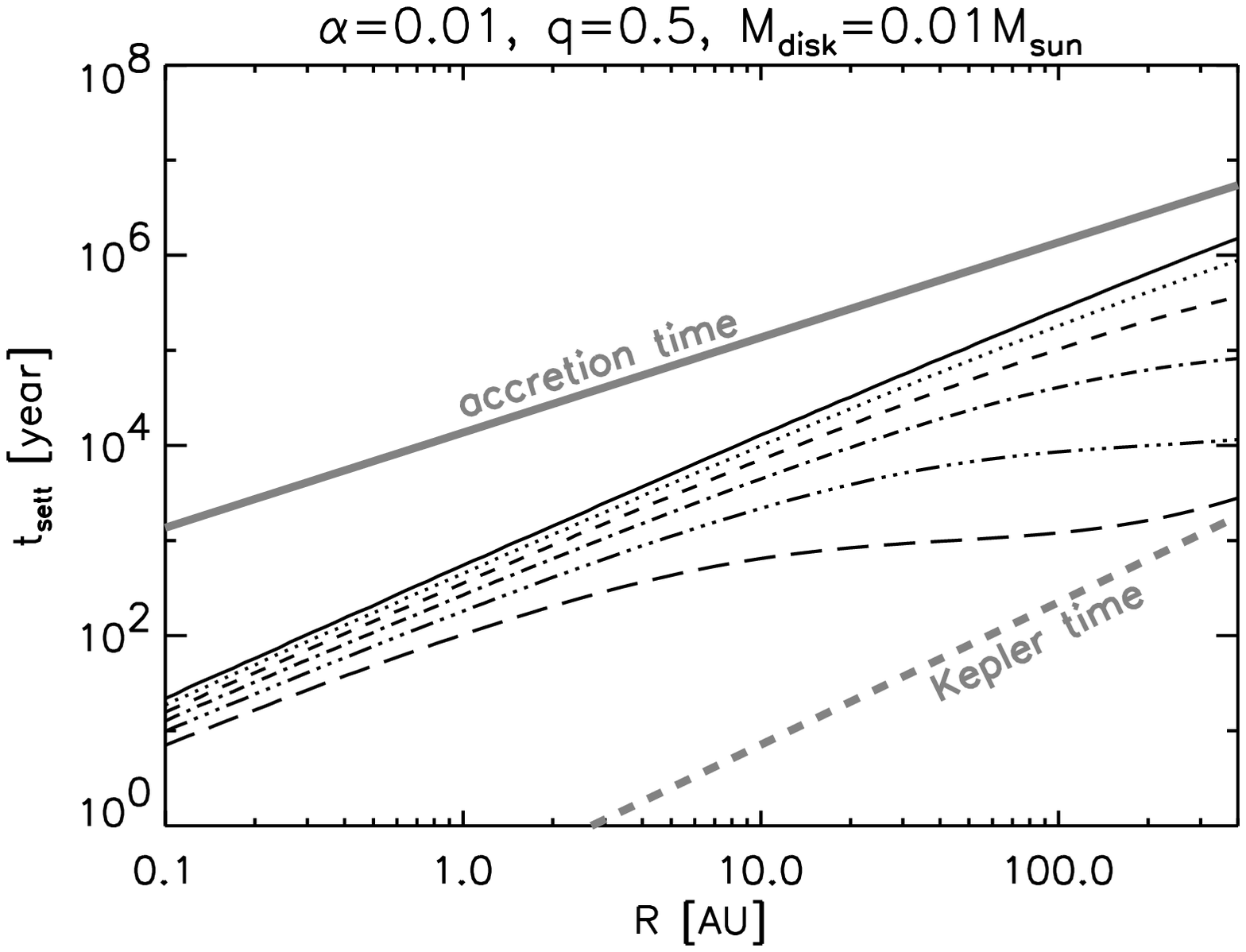}}
\caption{\label{fig-analytic-1}\emph{Left:}
the dust depletion height $\zdepl$ for the disk model described in the text,
for $\alpha=0.01$ and $q=1/2$. This is the height above which virtually all
grains of a certain size have been removed by settling, once the equilibrium
settling-stirring solution is reached.  The solid grey curve shows the
$\tau=1$ surface of the disk at $\lambda=0.55\mu$m if the dust has not yet
settled and the opacity is dominated by grains smaller than $0.1\mu$m.
\emph{Right:} the time scale $\tset$ for reaching such an
equilibrium. \revised{This is computed from Eq.(\ref{eq:10}) with $z=\zstir$
(i.e.~not with $z=\zdepl$, since we are interested here in finding the time
scale for the entire vertical equilibrium to set in).} The solid grey curve
is the viscous (i.e.~accretion) time scale $\taccr=R^2/\nu$. The dotted grey
curve is the Kepler time scale $\tkep=1/\Omega_K$.  Since the settling
cannot happen on a time scale smaller than this $\tkep$, we have added
$\tkep$ to the plotted time scale to ensure this minimum. The annotation for
the black curves is the same as in the left panel.}
\end{figure*}

\begin{figure*}
\centerline{
\includegraphics[width=9cm]{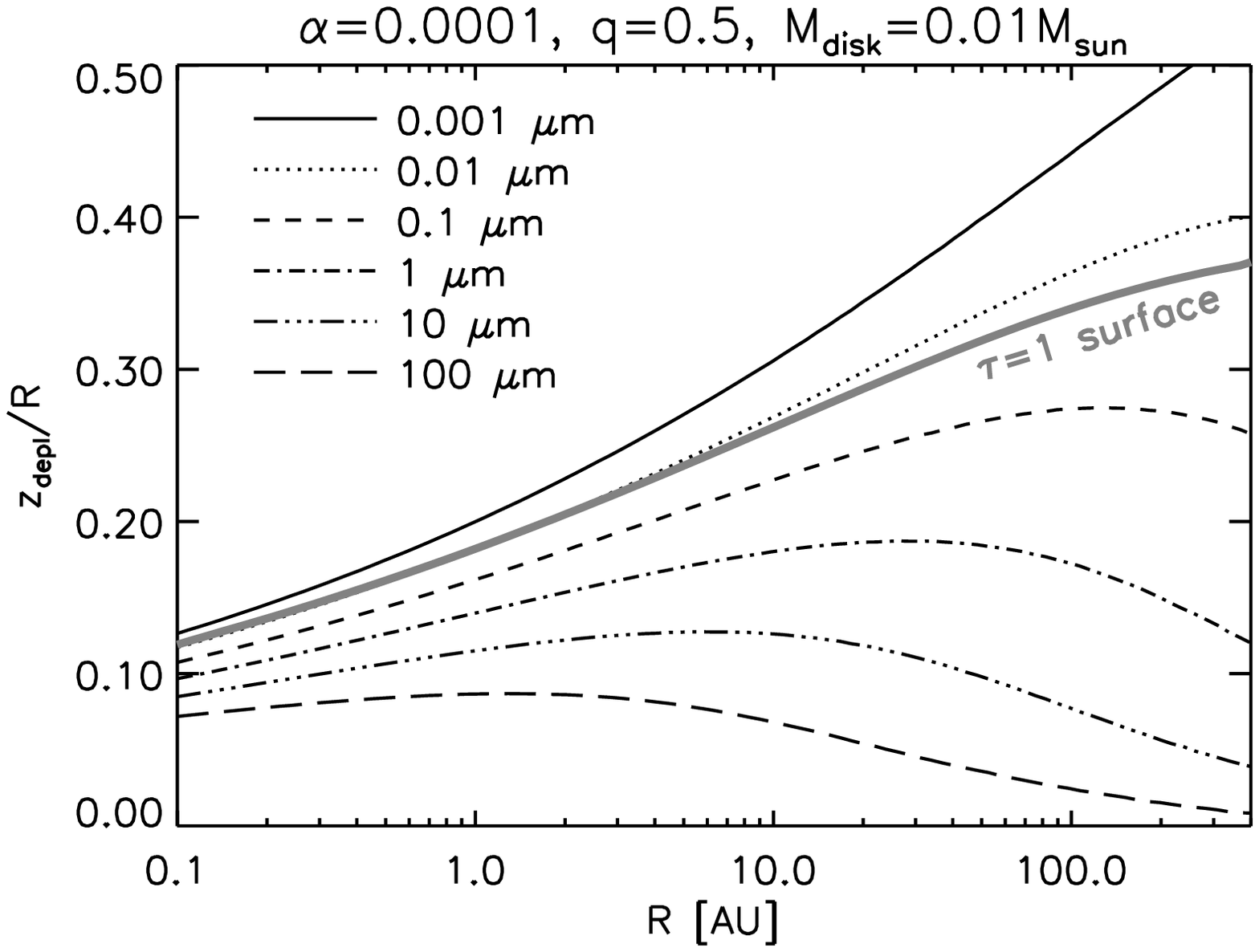}
\includegraphics[width=9cm]{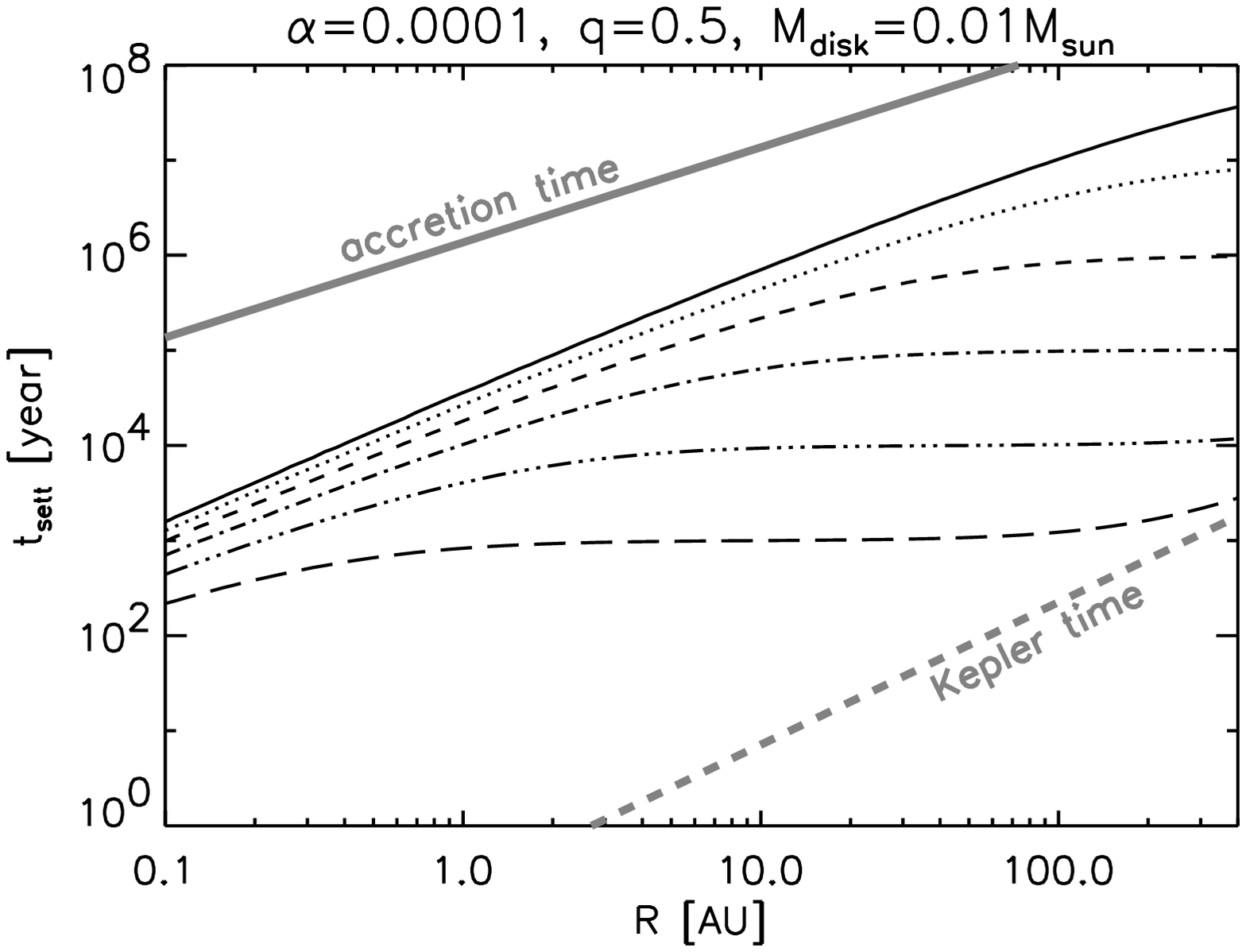}}
\caption{\label{fig-analytic-2}As Fig.~\ref{fig-analytic-1}, but now for
$\alpha=0.0001$.}
\end{figure*}

In Fig.~\ref{fig-analytic-1} the resulting 
$\zdepl/R$ and the time scale at which the equilibrium state is reached are
shown for the case of $\alpha=0.01$ and $q=1/2$. In
Fig.~\ref{fig-analytic-2} the same is shown, but this time for
$\alpha=0.0001$. Shown in the left panels in both figures is a grey line,
which indicates the initial height of the $\tau=1$ surface for radially
outward moving stellar photons before the grains have started to settle. It
is interesting to observe that for the $\alpha=0.0001$ case the height
$\zdepl/R$ below which the grain settle (even the 0.1$\mu$m grains) is below
this initial $\tau=1$ surface height. Since dust grains carry virtually all
of the opacity of the disk, this means that when the grains settle, the
photospheric surface height $\Hs$ of the disk is reduced, i.e.~the optical
appearance of the disk becomes flatter. Also the infrared emission from the
disk is reduced as the disk then captures less stellar flux.

The steady state is always reached on a shorter time scale than the
accretion time scale ($\taccr\equiv R^2/\nu$). It is shorter at small radii
than at larger radii, but beyond a certain radius it stays constant. This is
because these grains have settled below one pressure scale height, and for
$\Sigma\propto R^{-1.5}$ the time scale for a grain to settle down to one
pressure scale height or lower is independent of radius
\citep{miyakenaka:1995}.  The settling time scale for 0.1 $\mu$m grains at a
few hundred AU ($\sim 10^6$ year) is not much smaller than the typical life
time of protoplanetary disks ($\sim 10^6\cdots 10^7$ year).

An important feature of the steady state solutions is that the $\zdepl(R)/R$
does not always form a flaring shape. Beyond a certain radius $\rturn(a)$
(different for different grain sizes), the value of $d(\zdepl(R)/R)/dR$
becomes negative. \revised{If the disk has only one grain size, and assuming
that $\zdepl$ is a good estimate of the resulting surface height $\Hs$ of
the disk, this would imply that the disk has an $\Hs/R$ that is not a
monotonically increasing function of $R$. This is a strong indication that
{\em self-shadowing effects} play a role in such disks:} the outer disk has
a lower $\Hs/R$ than the intermediate regions and therefore resides in the
shadow of these intermediate disk regions \revised{(note that if one defines
$\Hs$ to be the location of the $\tau=1$ surface for radially-outward moving
photons, a lower $\Hs/R$ at larger radii is meaningless because of the
shadowing; see Fig.~\ref{fig-fulldisk-single-tausurf})}. For the case of
$\alpha=0.01$ and 0.1 $\mu$m grains this effect is small. One must go to 10
or 100 $\mu$m grains to see this effect. For $\alpha=0.0001$ this effect
already takes place with 0.1 $\mu$m grains. 

\revised{In the calculations shown above the Stokes number remains mostly
well below unity for grain sizes up to 100 micron. However, if we take
$q=0$, the Stokes number becomes significantly larger than unity beyond the
point where $d(\zdepl(R)/R)/dR$ becomes negative. This increases the
shadowing effect since the $\zdepl(R)$ drops more quickly as a function of
$R$ than for the case $q=1/2$. The radius $\rturn$ of this turn-over point,
however, does not change considerably.}

\begin{figure}
\centerline{\includegraphics[width=9cm]{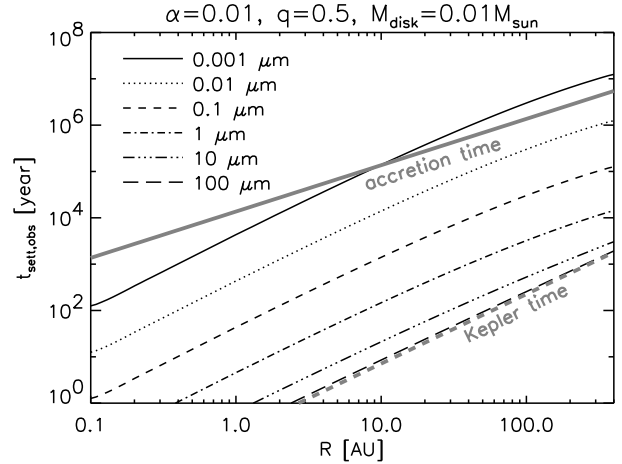}}
\caption{\label{fig-timescl-hide}The time scale with which grains settle
below the $\tau=1$ surface. The solid grey curve is the
viscous (accretion) time scale for $\alpha=0.01$. The dotted grey curve is
the Kepler time scale $\tkep=1/\Omega_K$. Since the settling cannot happen
on a time scale smaller than this $\tkep$, we have added $\tkep$ to the
plotted time scale to ensure this minimum.}
\end{figure}
It should be noted that the time scale for the settling can be deceptive.
It is the time scale it takes to reach the stationary state: the equilibrium
between turbulence and settling. But the very upper parts of the disk get
depleted from dust more quickly, since the dust settling velocity increases
with the decreasing density. The time scale for settling of $0.1\mu$m grains
in the very outer regions of the disk is therefore an upper limit of the
settling time: on a much shorter time the dust may settle down below the
$\tau=1$ surface of the disk, which means that settling can quickly affect
the shape of the photospheric surface of the disk. In
Fig.~\ref{fig-timescl-hide} the time scale is shown within which grains
settle below the $\tau=1$ surface. This $\tau=1$ surface is computed for
radially outward moving photons, using only the absorption opacity for
silicate grains of 0.1 $\mu$m size at $\lambda=0.55\mu$m
(i.e.~$\kappa=2\times 10^3\cm^2/\gram$), and is shown as the grey curve in
Figs.\ref{fig-analytic-1} and \ref{fig-analytic-2}.

\section{1-D vertical models: How the upper layers deplete}
\label{sec-vert-models}
The analytic estimates presented above give a feeling for the spatial
and time scales on which the dust settling takes place.  If we would
only be interested in the steady state solution, we could also use the
analytical expressions derived by Dubrulle et al
(\citeyear{dubmorster:1995}).  However, in order to see how an
initially fully mixed disk evolves through settling, we prefer to use
a numerical solution from which we can extract the time variations.
We do this in two steps.  First we take a close look at the local
solution at a given distance from the star.  In
section~\ref{sec:1+1-d-disk} we will then integrate over the entire
disk in order to study the emerging SED and disk images.

Using the expressions for the settling velocity and the turbulent diffusion
coefficient presented in section \ref{sec-estimates}, it is straightforward
to derive a one-dimensional vertical advection-diffusion equation for the
motion of grains. In this one-dimensional approximation we neglect radial
migration of the grains due to gas drag or other effects (Weidenschilling
\citeyear{weidenschilling:1977}). Also we ignore the accretional evolution
of the disk itself, or radial mixing phenomena. We only consider the
vertical motions of the grains induced by settling and vertical turbulent
stirring. We write the distribution function of grains of mass $m$ located
at height $z$ above the midplane as $f(m,z)$. It is defined such that
$f(m,z)\,dm\,dz$ represents the number of grains (of this size, at this
height) per square cm of disk.

\begin{figure*}
\centerline{\includegraphics[width=9cm]{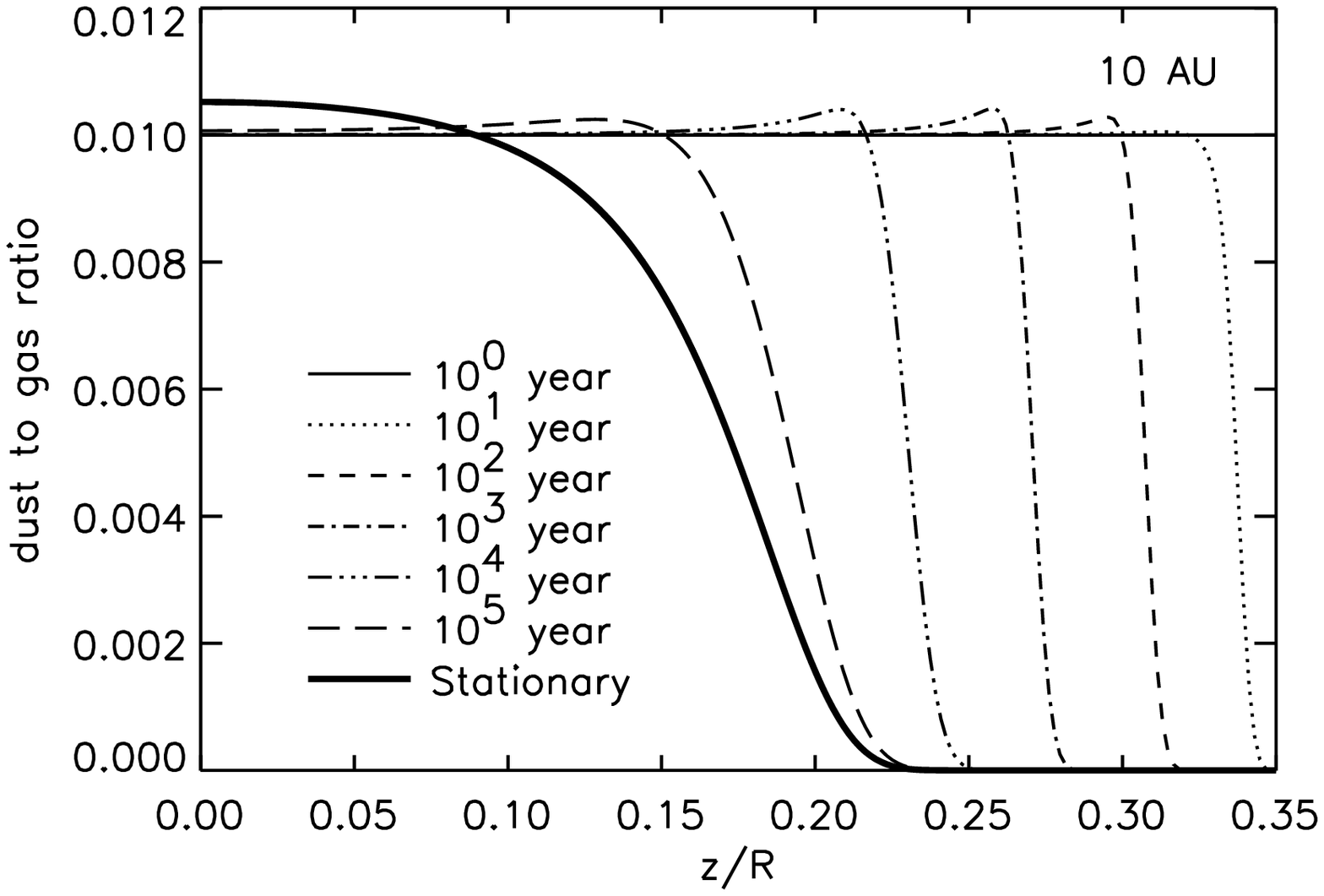}
\includegraphics[width=9cm]{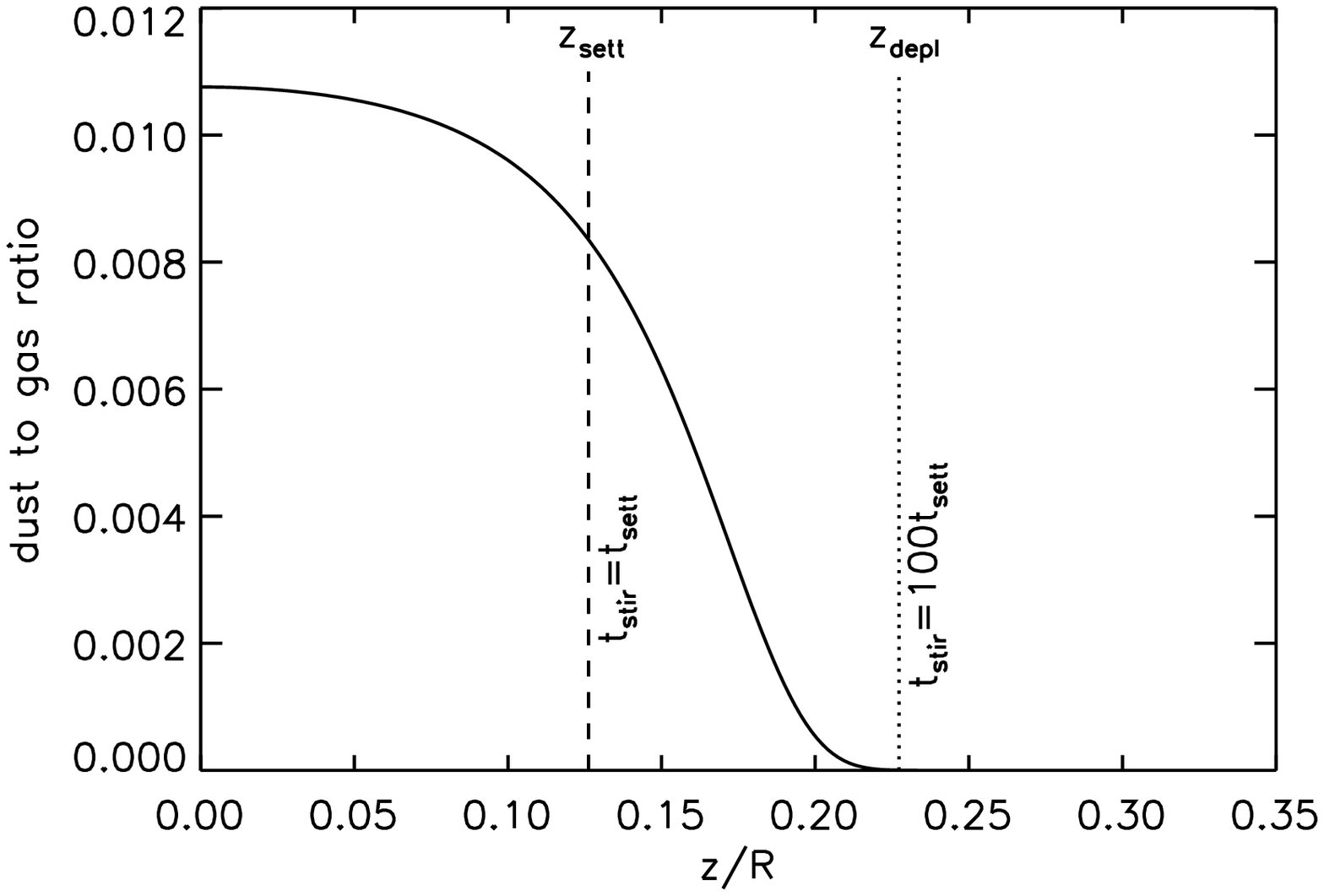}}
\caption{\label{fig-oneslice}The dust-to-gas ratio in a 1-D vertical slice
at 10 AU in the standard disk model \revised{with $\alpha=0.0001$}, assuming
$0.1\mu$m size grains. At the start ($t=0$) the dust is well-mixed with the
gas, i.e.~the dust-to-gas ratio is constant (we take it 0.01). Left: the
time sequence. Right: the stationary solution compared to the analytic
estimates. The vertical lines are the computed values of $\zstir$ and
$\zdepl$ (from left to right). These are defined to be solutions of
Eq.(\ref{eq-zsett}) with $\fact=1$ and $\fact=100$ respectively. As one can
see, $\zdepl$ is a good definition of the height above which the depletion
is almost perfect.}
\end{figure*}

For the gas we use again a Gaussian density profile and isothermal
temperature profile. We assume that during the dust settling
process this gas density and temperature profile do not change with time.

The conservation equation for the dust grains is (ignoring the $R$
coordinate for compactness):
\begin{equation}
\label{eq-settdiff-conserveq}
\begin{split}
\frac{\partial f(m,z)}{\partial t}
& -\frac{\partial}{\partial z}
\left[\rho_{\mathrm{gas}}D(m,z)\frac{\partial}{\partial z}
\left(\frac{f(m,z)}{\rho_{\mathrm{gas}}}\right)\right] +\\
& \frac{\partial}{\partial z}\left(f(m,z)v_{\mathrm{sett}}(m,z)\right)
=0
\fullstop
\end{split}
\end{equation}
This equation is valid if no dust coagulation takes place. 

We take the same stellar parameters as in Section \ref{sec-estimates}, and
the same disk setup. We take our slice at $R=10\AU$, which means that the
gas surface density at that point is $\Sigma=13\gram/\cm^2$. We simulate the
settling of 0.1 $\mu$m grains and adopt $\alpha=0.0001$ for the turbulence.
In Fig.~\ref{fig-oneslice} the evolution of the dust settling is shown for
this slice, and a comparison of the stationary solution (which is reached at
about $t=3\times 10^5\yr$) to the analytic estimates is given as well.

One can see that as the grains settle, the upper parts of the disk are
entirely depleted from dust. After only $10^3$ years the depletion above
$z/R=0.30$ is a factor of $10^{-10}$ and above $z/R=0.325$ it is $10^{-30}$.
This means that long before a steady state is reached, the regions above
$z/R=0.30$ are already completely dust-free. One can see in
Fig.~\ref{fig-oneslice}-left that the dust that has settled forms a
layer on top of the disk in which the dust-to-gas ratio is slightly above
0.01. This is simply the piling-up of the settled dust from higher region of
the disk.

It is interesting also to see that while the dust is still settling it
has a very sharply defined surface. Below a certain $z$ the
dust-to-gas ratio is about 0.01, and just a small step above this $z$
the dust-to-gas ratio is almost zero.  This sharp transition is a
consequence of the fact that initially all grains are settling with
the equilibrium settling velocity, unperturbed by stirring.  Since the
settling speed decreases towards the mid-plane, the upper-most grains
catch up with grains just below them and therefore produce a very sharp
boundary.  In the later stages, when the steady state has been
reached, this sharp cut-off softens, but it should still be
noticed that beyond $z\simeq 0.23R$ the depletion increases very fast.

The $\tau=1$ surface of the disk, i.e.~the disk's photosphere, for silicate
grains of 0.1 $\mu$m at $t=0$ (i.e.~the well-mixed case without settling)
lies in this example at about $\Hs/R=0.3$. 
%
%
Note that this is the $\tau=1$ surface for the stellar radiation which
enters the disk under a grazing angle of $\phi=0.05$.  After $10^3$ years
the dust has settled well below this point, meaning that the dust will take
its $\tau=1$ surface along with it toward lower $z$ (as already predicted in
Section \ref{sec-estimates}). In other words: after $10^3$ years the
photospheric surface height of the disk $\Hs$ is reduced as a result of dust
settling. The gas still remains at higher $z$, but it is the dust that
carries the opacity and determines the disk's appearance.

\section{1+1-D disk models: SEDs and images}\label{sec:1+1-d-disk}
As we have seen above, dust settling can strongly affect the shape of
the $\tau=1$ surface $\Hs(R)$ of the disk. This shape determines for a
large part the thermal infrared emission from this disk and also the
appearance of images in scattered light.  If the disk's surface has a
strongly flaring shape, one expects a spectral energy distribution
(SED) with a strong emission, particularly at mid- and far-IR
wavelengths.  If the disk has such a shape that the outer regions are
shadowed by the inner regions (if $\Hs/R$ decreases with radius), then
the expected SED has a rather weak far-IR excess. The predicted images
in scattered light for these two cases are also expected to be very
different. It is therefore of crucial importance to investigate how
the shape of the disk changes as a result of dust settling.

We again start with the simple model described earlier, and we do a 1-D dust
settling calculation at every radius, taking a radial grid of 20 grid points
in $R$. In this way we simulate the entire disk in a \mbox{1+1-D} manner.
\revised{We take $\alpha=0.00001$, i.e.~a factor of ten smaller than taken
in Section \ref{sec-analytic-est}, in order to accentuate the effects of
settling on the appearance of the disk}.  At $t=$1, 10, \ldots, 10$^6$ years
we apply a 2-D radiative transfer code (called {\tt
RADMC}\footnote{www.mpa-garching.mpg.de/PUBLICATIONS/DATA/ radtrans/radmc/})
to the given dust distribution, and compute the SED of the disk at these
times.  Also we compute the intensity of scattered light as a function of
distance from the star, i.e.~the brightness of an images of the disk in
scattered light at $\lambda=1.65\mu$m.

The dust temperature computed by {\tt RADMC} is likely to differ
somewhat from the dust/gas temperature $T(R)$ assumed as input to
the density structure of the disk.  In principle one should iterate the
computation of the temperature and the density structure to obtain a
self-consistent model as we did in DD04.  However, since our main
interest is the process of dust settling rather than the details of
disk structure, we ignore this feedback for the sake of simplicity.

\subsection{Model with single grain size}
\begin{figure*}
\centerline{
\includegraphics[width=9cm]{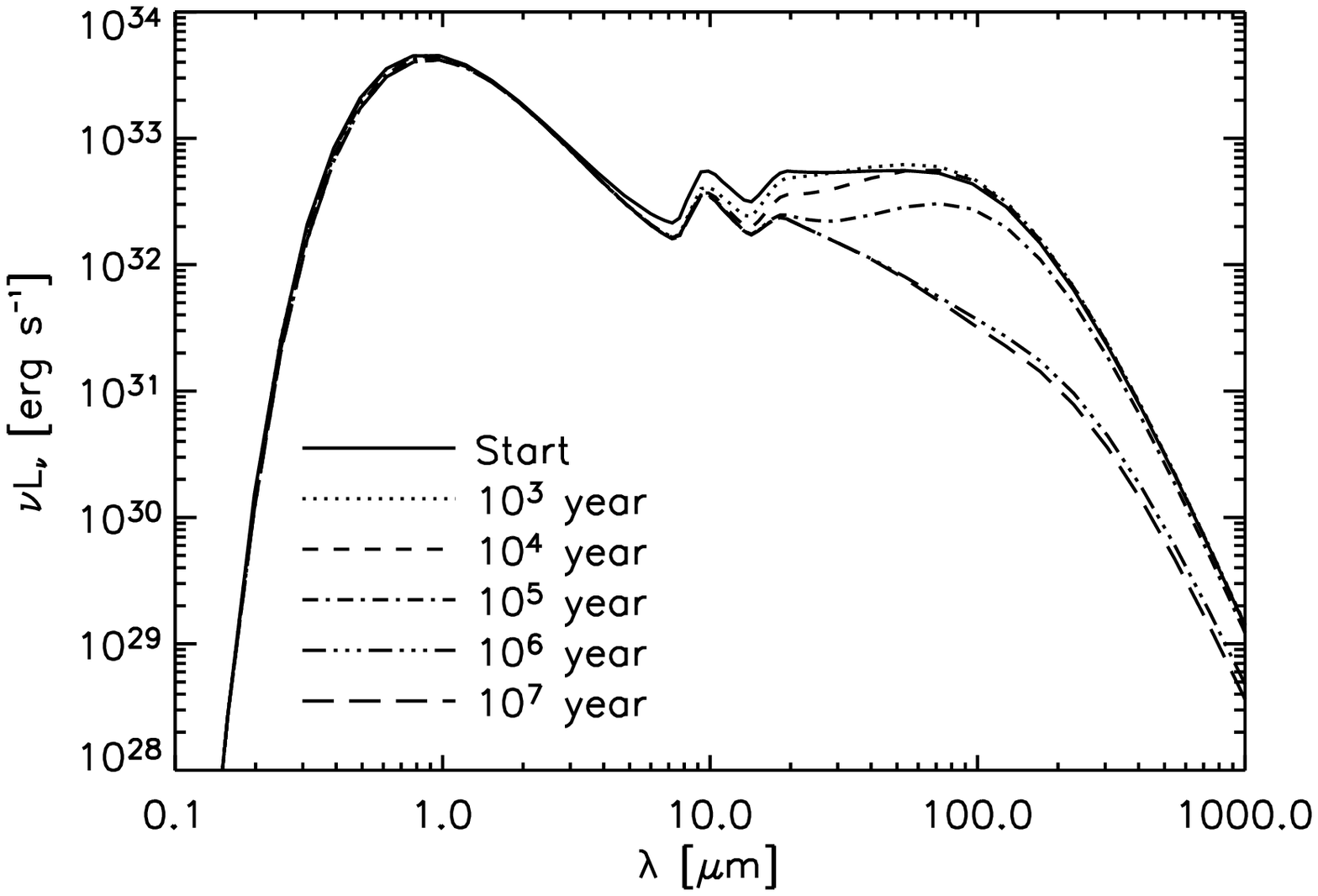}
\includegraphics[width=9cm]{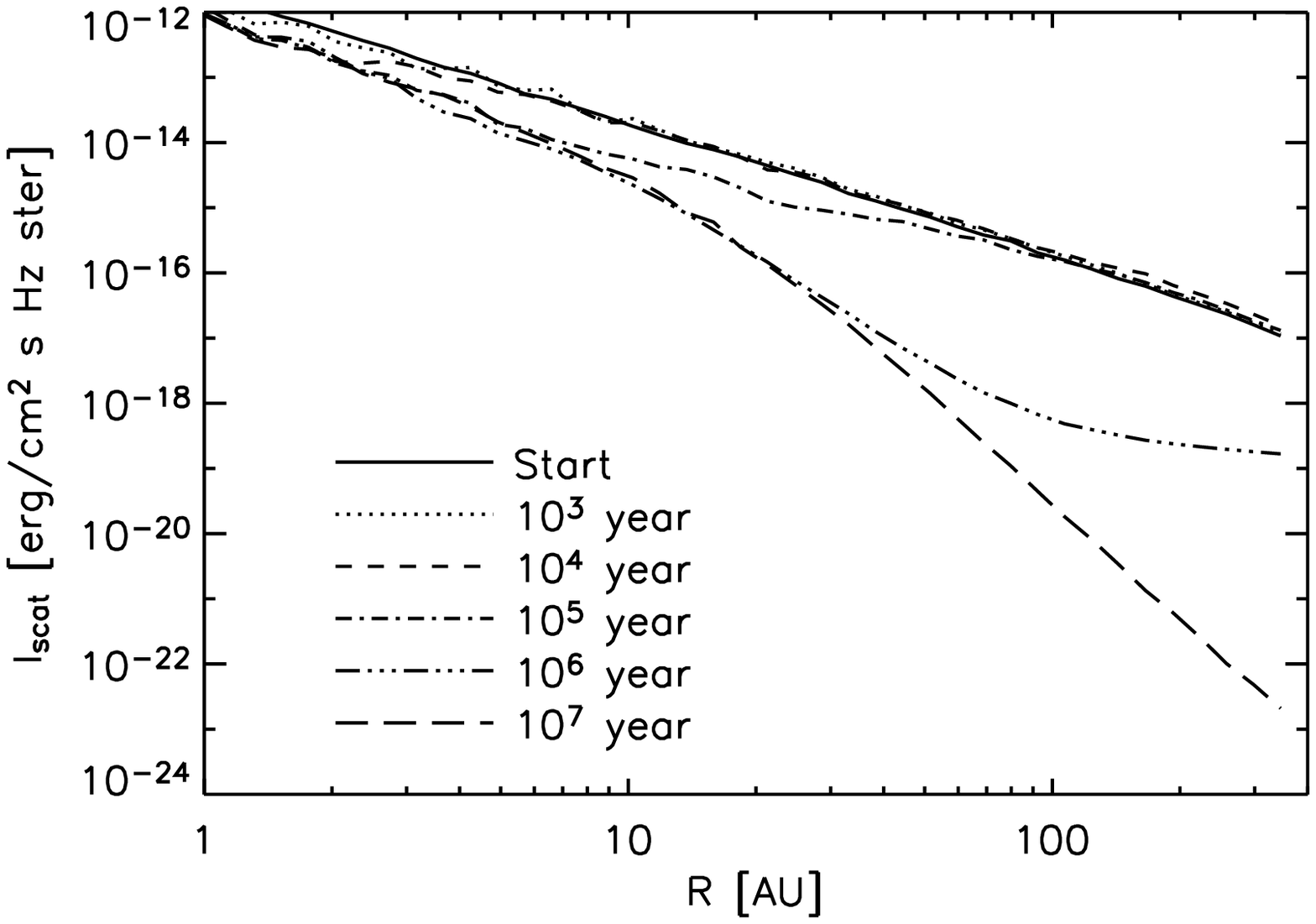}
}\caption{\label{fig-fulldisk-single-time}The SED (left) and the $0.55\mu$m
scattering intensity (right) of the disk as a function of time for the 1+1D
model with a single grain size of $a=0.1\mu$m. Note that the time intervals
are logarithmic (a factor of 10 apart). \revised{The disk is seen face-on.}}
\end{figure*}

The first model (model A) only has grains of $0.1\mu$m size. In
Fig.~\ref{fig-fulldisk-single-time}-left the SED as a function of time is
shown.  The SED changes only barely on time scales smaller than $10^5$
years.  But on time scales longer than $10^5$ years the far-IR flux drops.
Two effects play a role in reducing the far-IR flux.  First of all, due to
the settling, the surface height of the disk $\Hs$ has been reduced.  Since
the thickness of a disk is directly related to the total amount of stellar
radiation captured and reprocessed by the disk, this effect reduces the
integrated IR emission over the entire wavelength domain.  Chiang et
al.~(\citeyear{chiangjoung:2001}) used this fact to argue that dust settling
may be responsible for the low far-IR flux of some Herbig Ae/Be star.  The
other, more important reason for the reduction of the far-IR flux is that
due to the turn-over in the $\Hs/R$ beyond $R=\rturn$, the outer disk is
\emph{shadowed} by the inner regions of the disk.  This means that these
outer regions are irradiated only in an indirect way, for example by light
that scattered off dust grains hovering slightly above the photosphere at
the turn-over (i.e.~self-shadowing) radius $\rturn$, or by radial radiative
diffusion (see DD04 for a discussion of self-shadowed disks).

\begin{figure}
\centerline{
\includegraphics[width=9cm]{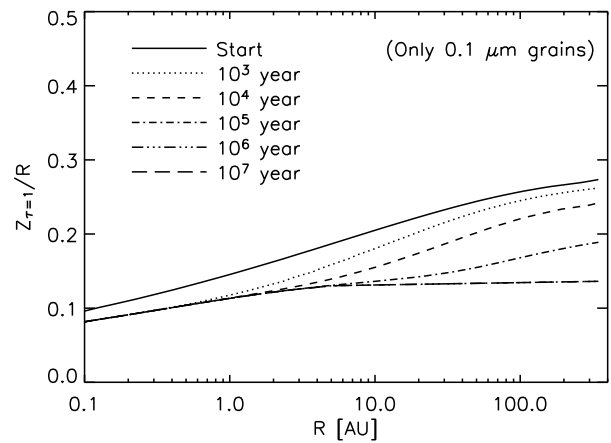}}
\caption{\label{fig-fulldisk-single-tausurf}The shape of the $\tau=1$
surface, where the optical depth $\tau$ is at 0.55$\mu$m and is measured
{\em radially} from the stellar surface outward (i.e.~horizontally in the
figure). For this reason the $\tau=1$ surface is a monotonically increasing
function of $R$, even when the disk fully collapses beyond some point. This
is indeed apparent in this figure, because it appears if beyond $10^6$ and
$10^7$ years nothing changes, whereas the scattering image still becomes
dimmer in this time frame (Fig.~\ref{fig-fulldisk-single-time}).}
\end{figure}

The time evolution of the scattering intensity shows a very interesting
behavior, as is shown in Fig.~\ref{fig-fulldisk-single-time}-right.
Initially, the scattering intensity changes only in the innermost regions
(inward of 10 AU), from the inside out.  After about 10$^5$ years, the inner
disk stabilizes - the steady state solution is reached there. By this time
the outer part of the disk up to 100 AU starts to be affected by the
settling. After 10$^6$years, the scattering intensity is already reduced by
more than a factor of 100, and after 10$^7$years in the outermost regions,
the reduction is even 10$^6$. This is clear evidence of the self-shadowing
effect operating in these regions.  Without shadowing, a much more gradual
change like the one seen in the inner regions would occur also here.  A
self-consistent computation of the disk structure would only enhance this
effect since the temperature in the outer disk will drop and lead to a
reduced disk scale height. In Fig.~\ref{fig-fulldisk-single-tausurf} the
self-shadowing effect is shown by the shape of the $\tau=1$ surface for
radially outward moving stellar photons. Up to $10^5$ years the $z/R$
location of the $\tau=1$ surface still increases for increasing $R$, which
is indicative of a flaring geometry. But after $10^6$ years the $\tau=1$
surface stays essentially at constant $z/R$ beyond about 7 AU, which 
indicates that the disk is shadowed beyond 7 AU.

\subsection{Model with grain size distribution}
In section~\ref{sec-estimates} we have shown that the settling behavior of
grains varies strongly with grain size.  It may be possible to represent the
grain size distribution with a typical size for a specific application, but
the correct size to use will depend upon the application. The intensities
seen in scattered light is dominated by one grain size, the infrared
emission by another grains size, and other indicators like PAH emission will
depend on yet another grain type. Therefore we now show the results of a
calculation in which we include a full size distribution. The grains of each
size settle with their own settling velocity and have their own Schmidt
number. The bigger grains will settle much faster and deeper than the
smaller ones, so that one obtains a different grain size distribution at
different heights above the midplane. As our initial grain size distribution
we take an MRN distribution (Mathis, Rumpl \& Nordsieck \citeyear{mrn:1977})
between grain sizes of 0.005 $\mu$m and 0.25 $\mu$m. The original MRN
distribution is only well constrained down to grain sizes of about 0.02
$\mu$m \citep{1993AAS...182.0813K}.  However, in order to see what the
influence of smaller grains would be we extend the size distribution down to
0.005 $\mu$m.

\begin{figure*}
\centerline{
\includegraphics[width=9cm]{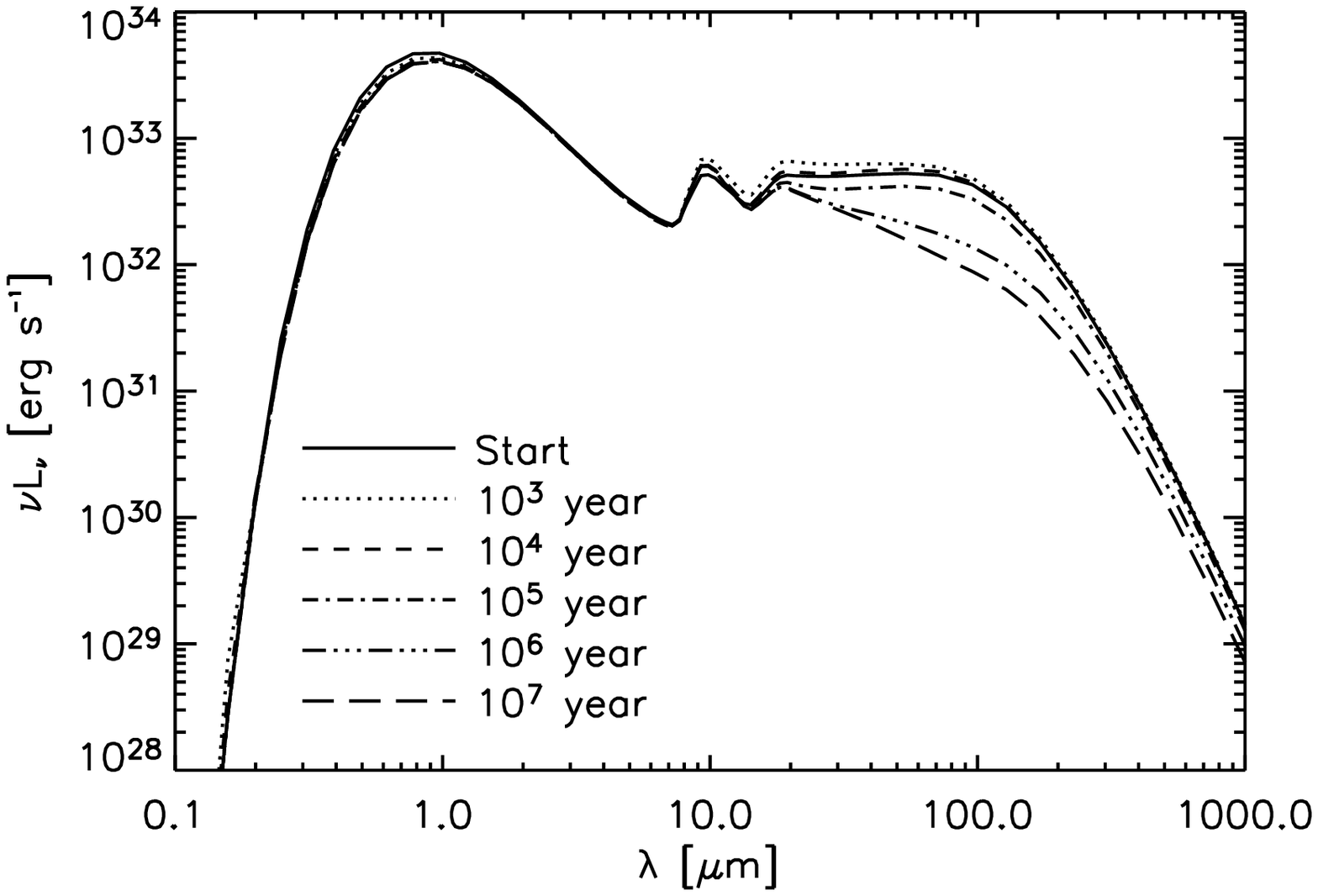}
\includegraphics[width=9cm]{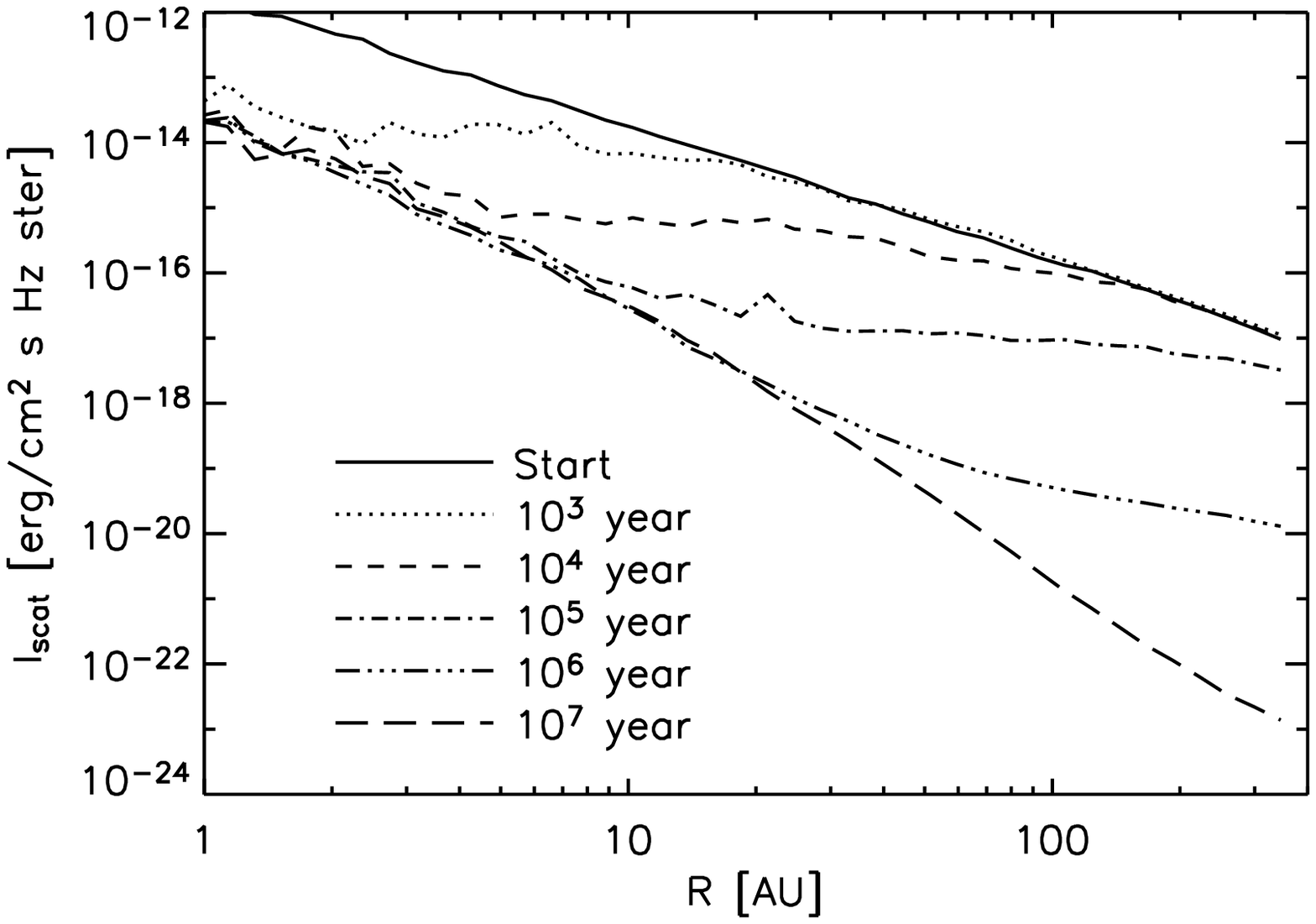}}
\caption{\label{fig-fulldisk-mrn-time}As
Fig.~\ref{fig-fulldisk-single-time}, but for the disk model with 
an MRN grain size distribution.}
\end{figure*}

The results are shown in Fig.~\ref{fig-fulldisk-mrn-time}. The same
shadowing effects are seen as in the case of the single $0.1\mu$m size
grains, although there are also some differences. It is interesting that
this shadowing in fact takes place also for the MRN distribution, even
though the very smallest grains clearly stay afloat much higher up. However,
these small grains have relatively low opacity at the stellar wavelengths:
the opacity is highest for grains between $0.1\mu$m and $1\mu$m.  Moreover
the MRN distribution goes as $m^2 f(m)\propto m^{1/6} \propto a^{1/2}$,
meaning that most of the mass is concentrated in the bigger grains. So even
if the opacity of the grains (per gram dust) would be independent of grain
size (up to 1 $\mu$m), still the opacity will be dominated by the $0.1\cdots
1\mu$m grains.  Therefore the very small grains will not account for much of
the reprocessing of stellar radiation into the infrared.  Since the albedo
of the small grains at stellar wavelengths is virtually zero, they will not
contribute at all to the scattering of starlight into the line of sight. In
effect, only the grains larger than roughly $0.03\mu$m will contribute
significantly to the reprocessing and scattering. Therefore, even though the
very small grains virtually do not settle at all, the {\em relevant} grains
($a\gtrsim 0.03\mu$m) {\em do} settle.

\begin{figure}
\centerline{
\includegraphics[width=9cm]{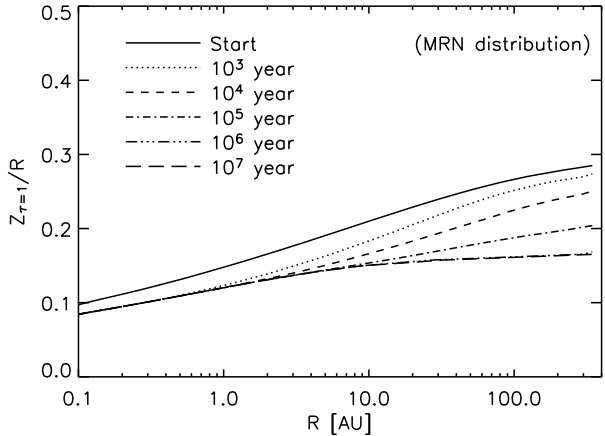}
}\caption{\label{fig-fulldisk-mrn-tausurf}
As Fig.~\ref{fig-fulldisk-single-tausurf}, but for the MRN model.}
\end{figure}

This has two effects, both of which contribute to the dimming of the
scattering image. First, the large grains which are most reflective fall
below the $\tau=1$ surface of the small grains (which are not
reflective). The small grains may therefore partly shield the larger
(reflective) grains from stellar light, thereby dimming the scattering
image. This `differentiation' effect is clearly seen in
Fig.~\ref{fig-fulldisk-mrn-time} as an overall dimming of the disk even in
regions which are clearly non-shadowed. \revised{This dimming is quite
strong in this case, but for higher values of $\alpha$ this effect is
reduced. As expected,} such an overall dimming is absent in the
single-grain-size case (Fig.~\ref{fig-fulldisk-single-time}). Secondly, the
$\tau=1$ surface itself is dragged down by settling, creating a
self-shadowed geometry similar to the one for the single grain size
simulation. The fact that even in this MRN simulation the disk becomes
partly self-shadowed can be seen from Fig.~\ref{fig-fulldisk-mrn-tausurf},
where the evolution of the $\tau=1$ surface is shown for the MRN
model. \revised{But it can also be seen from this figure that the shadowing
is not as sharp as in the case of a single 0.1 $\mu$m grain. Indeed, the
weakening of the far-infrared flux for the MRN case is less pronounced than
in the single-grain model. For the scattered light intensity the effect of
self-shadowing is still clearly seen, though the dimming by the
`differentiation' effect may, for these low values of $\alpha$, be equally
important for the (non-)detection of disks in scattered light.}

In Fig.~\ref{fig-fulldisk-mrn-time}-left the evolution of the SED is shown
for the MRN model. Interestingly, the mid- to far-infrared excess first
slightly brightens (between 0 and $10^4$ years). Only afterwards the mid- to
far-infrared excess starts to decline, eventually becoming rather weak at
$10^6\cdots 10^7$ years. This initial brightening is because due to the quick
settling in the inner region, while the outer regions take a longer time,
the flaring angle of the surface of the disk temporarily increases. Only
after the outer regions start to sink into the shadow of the intermediate
disk regions the mid- to far-infrared excess starts to decrease
significantly.

\begin{figure}
\centerline{
\includegraphics[width=9cm]{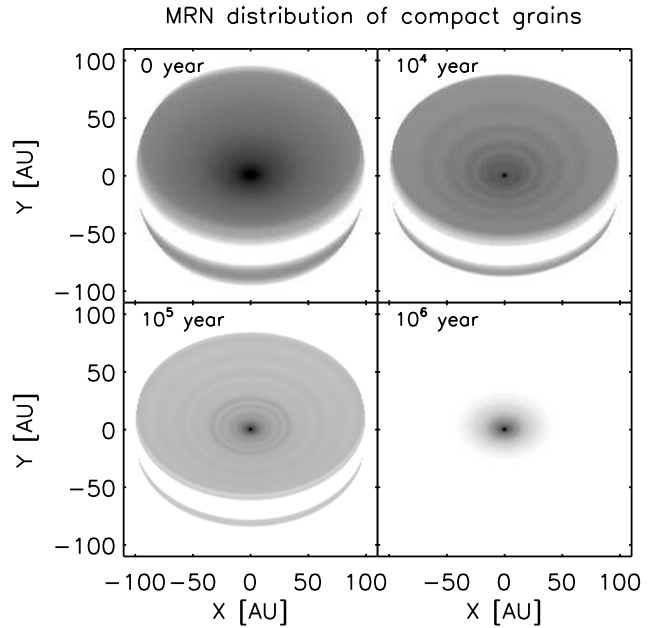}}
\caption{\label{fig-fulldisk-mrn-images}\revised{Images of the
$\alpha=0.00001$ disk model with an MRN distribution of compact grains (see
Fig.~\ref{fig-fulldisk-mrn-time}), seen at an inclination of $45^{o}$ at
$\lambda=0.55\mu$m. For these images the model was truncated at 100 AU, so
that both the inner regions ($\simeq 25$ AU) and the outer regions ($\approx
100$ AU) can be seen clearly. The images are at 0, $10^4$, $10^5$ and $10^6$
years. The gray scale is logarithmic intensity (inverse video). Note that
the central star is not included in these images (a perfect coronograph).}}
\end{figure}

\revised{Fig.~\ref{fig-fulldisk-mrn-images} shows images of the disk at 0.55
$\mu$m at inclination 45$^{o}$. We truncated the disk to 100 AU so that both
the inner regions and the outer regions can be seen clearly. }

\section{Discussion}

\subsection{Scaling of the results}
The analytic formulae of Section \ref{sec-estimates} and the numerical
models of Section \ref{sec-vert-models} can easily be scaled to the case of
fluffy grains. The critical parameter that goes into all these equations is
the ratio of grain geometric cross section and grain mass: $\sigma/m$. A
compact silicate grain with a specific weight of $\chi=3.6$ g/cm$^3$ and a
size of $a=0.1\mu$m has $\sigma/m=3/(4a\chi)=2\times 10^4$ cm$^2$/g. A fluffy
grain of size $10\mu$m but the same value of $\sigma/m=2\times 10^4$ cm$^2$/g
will behave exactly the same as the compact 0.1 $\mu$m size grain.

\subsection{The role of viscous dissipation}\label{sec-visc-diss}
\revised{Throughout the paper we have taken a very simple temperature and
density structure of our disk. In reality the structure of an irradiated
disk will be more complex (see e.g.~Dullemond, van Zadelhoff \& Natta
\citeyear{dulvzadnat:2002}; DD04). Moreover, if the viscous heating by
active accretion is included, the midplane can become considerably hotter
than in our simple model. While a detailed inclusion of these effects
is beyond the scope of this paper, it is instructive to make some
estimates.}

\revised{The midplane temperature for a disk heated only by viscous
dissipation is given by $T_{\mathrm{v,mid}}^3=
(27/64)(\alpha\kappa_{R}k/\sigma\mu m_p)\Sigma^2\Omega_K$, and the surface
temperature (``effective temperature'') is given by
$T_{\mathrm{v,eff}}=(8/3\kappa_R\Sigma)^{1/4}\,T_{\mathrm{v,mid}}$, where
$\kappa_R$ is the Rosseland mean opacity of the dust+gas mixture (before
settling). By finding the radius beyond which the $T_{\mathrm{v,mid}}$ and
$T_{\mathrm{v,eff}}$ fall below the temperature given by
Eq.(\ref{eq-temp-irrad}), we can estimate beyond which radii the viscous
heating becomes unimportant at the midplane and the surface respectively.
For the model parameters given in Subsection \ref{sec-analytic-est}, for the
case of $\alpha=0.01$ these radii are 5 AU and 0.6 AU, for $\alpha=0.0001$
they are 1 AU and 0.05 AU, and finally for the case of $\alpha=0.00001$
these are 0.5 AU and 0.02 AU respectively. At large radii the viscous
heating is therefore not important, but it may be important at small
radii.}

\revised{An increase in temperature due to viscous heating will increase the
friction of the particles, and it will also increase the pressure scale
height of the disk. These combined effects will increase the height of the
photosphere after a settling equilibrium has set in. Since viscous heating
is unimportant at large radii, and therefore the increase in height of the
disk takes place predominantly at small radii, the effect of inclusion of
viscous dissipation is likely to increase the shadowing effect.}

\subsection{Geometries}
In our earlier paper (DD04) we have demonstrated, on the basis of
theoretical models, that disks around Herbig Ae/Be stars come in two main
shapes: flared disks and self-shadowed disks. In the self-shadowed case, the
inner rim of the disk casts a shadow over the entire disk behind it. The
cause for the self-shadowing was a reduction of opacity due to grain growth.

\begin{figure*}
\centerline{
\parbox[t]{7.2cm}{
\centerline{\parbox[b]{6.5cm}{\mbox{}\vspace{0.0cm}\\
\includegraphics[width=6.5cm]{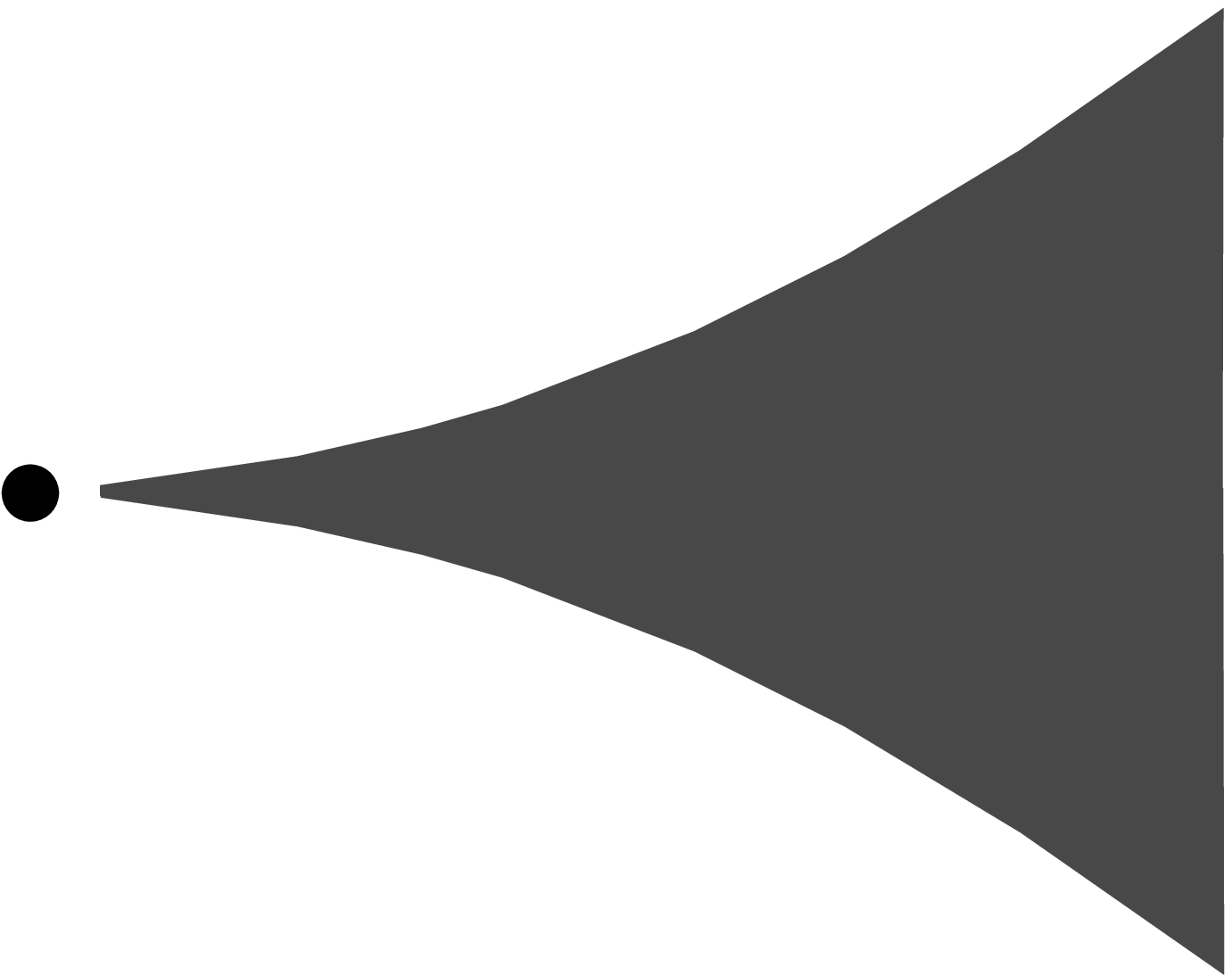}
\\\vspace{0.0cm}\mbox{}}}
}\hspace{2em}
\parbox[t]{7.2cm}{
\centerline{\parbox[b]{6.5cm}{\mbox{}\vspace{1.1cm}\\
\includegraphics[width=6.5cm]{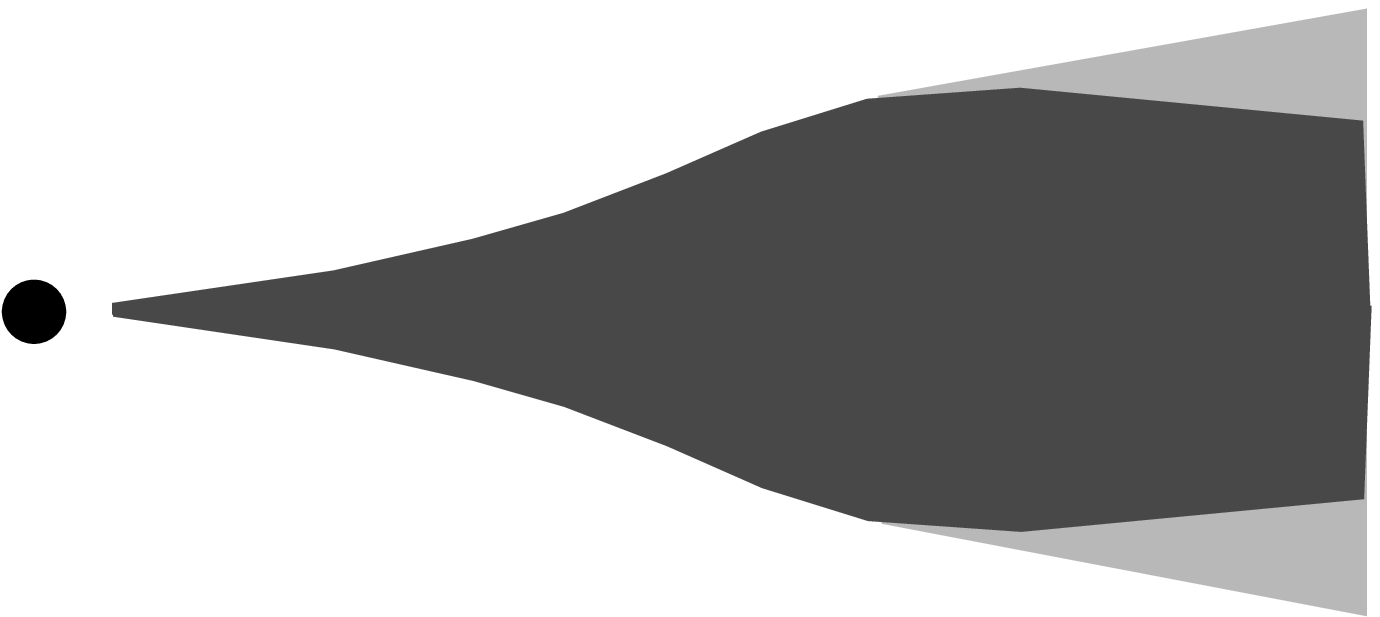}
\\\vspace{1.1cm}\mbox{}}}
}}
\caption{\label{fig-pictograms}Pictograms of the two geometries. Left:
the flaring disk we start out with. Right: after settling the outer 
parts of the disk are self-shadowed. Note that whether this
partly self-shadowed geometry is obtained depends whether the turbulence
is low enough.}
\end{figure*}

In the present paper we consider another process: dust settling. We ignore
grain growth. Moreover, we focus mainly on T Tauri stars, in which the
effect of the puffed-up inner rim is likely to be less pronounced, and we
ignored it here.  In this setup we find geometries very similar, though not
identical to DD04. We start out with a flaring disk, and after settling we
may obtain a partly self-shadowed disk, dependent on the value of
$\alpha$. In contrast to the self-shadowed disk solutions of DD04 our
self-shadowed solution are only self-shadowed outside of a few AU, while
they remain flaring inside of this radius. And the shadow is cast by the
flaring inner part of the disk instead of the puffed-up inner rim. In
Fig.~\ref{fig-pictograms} the two geometries are shown.

For Herbig Ae/Be stars the inner rim cannot be ignored that easily. It may
happen that dust settling could cause the entire disk to fall into the
shadow of the inner rim, producing the fully-self-shadowed geometry
presented in DD04, but in a different way than was shown by DD04.  However,
dust settling also affects the puffed-up inner rim, and this is more
difficult to model in detail. The inner rim is much hotter than the rest of
the disk.  A higher abundance of free electrons would cause the
magneto-rotational instability to operate more efficiently here and
increase turbulent stirring.  Including the inner rim into the analysis is
therefore a subtle matter, which we defer to a future paper.

\subsection{SED type versus scattering images}
In DD04 we showed that for Herbig Ae stars, disks dominated by small grains
usually have flared geometries, while disks in which most of the small
grains have coagulated to big grains are usually self-shadowed.  The SEDs of
these two kinds of disks were very reminiscent of the two types of SEDs
found among the SEDs of Herbig Ae/Be stars (the `group I' and `group II'
sources of Meeus et al.~\citeyear{meeuswatersbouw:2001}). It was therefore
suggested that group I sources are flared disks and group II sources are
self-shadowed disks.  This suggestion has passed several observational
tests, including the identification of UX Orionis stars with group II
sources (Dullemond et al.~\citeyear{dulancackboe:2003}), the presence or
absence of PAH emission features (Acke et al.~A\&A subm.), and the
correlation of the group I/II identification with the (sub-)mm slope of the
source (Acke et al.~subm.).  The self-shadowed solutions in DD04 were
caused by a simple optical depth effect: when most of the small grains are
turned into a few big ones, the optical depth is greatly reduced.  The disk
then gradually becomes self-shadowed, since at large radii the optical depth
is not sufficient to keep its surface above the shadow of the inner rim.  In
the current study we now show that also \emph{dust settling} may turn a
flared disk into a self-shadowed one.  In reality, it is most likely a
combination of \emph{settling and coagulation} leads to the observed
effects.

It was mentioned in DD04 that self-shadowed disks are expected to have weak
scattering images, for the obvious reasons that direct stellar radiation
cannot irradiate the disk. But it was also shown that only for extreme
parameters the self-shadowed disk solutions that follow from well-mixed
models are very deeply shadowed.  We estimate that for a fully mixed model,
the scattering intensity of the disk is reduced by typically a factor of
10. In the present paper we now show that dust settling can cause a much
stronger shadowing effect. It can remove all grains capable of scattering
stellar radiation (i.e.~grains larger than about 0.03 $\mu$m size) from the
upper layers of the disk, so that if self-shadowing occurs in this way, the
expected scattering images are extremely weak. Therefore, group II sources
have weaker scattering images than group I sources. If, on top of that, it
is found that the group II sources are {\em orders of magnitude} weaker,
then it is clear that dust settling must also play a role.

\subsection{Timescales}
The most realistic simulation in our study is the run involving an MRN grain
size distribution. As shown in figures~\ref{fig-fulldisk-mrn-time} and
\ref{fig-fulldisk-mrn-tausurf}, the changes in the SED and in the scattering
image brightness happen largely on a timescale of 10$^{6}$ years.  Given the
fact that the observed T Tauri stars are typically between 1 and 10Myrs old,
this gives an indication that settling might be happening too fast in our
simulation. If indeed the settling has happened already after 10$^6$ years,
one would expect most disks to show steep SEDs and weak scattering images.
Looking at the sample of Herbig stars studied by
\citet{meeuswatersbouw:2001}, this seems to be consistent in the way that
the group II sources\footnote{51 Oph is a special case of a star with an
extremely low IR excess and is therefore not considered here.} have ages
larger than 1 Myr while two of the group I sources with available ages are
only about 10$^5$ years old.  However, with AB Aur (about 2 Myrs) and in
particular HD 100546 (about 10Myrs), there are two stars in the flaring disk
group I which are significantly older.  For T Tauri stars, the situation is
less clear as the SEDs of T Tauri stars have not yet been classified in the
same way as the Herbig stars so far. There may be observational biases
involved here: The stars have been selected on IRAS colors which favors the
detection of group I sources.  If there is a large number of sources with
strongly flaring disks at ages much larger than 10$^6$ years, it is possible
that the dust settling study presented here is incomplete.  A possible
solution could be larger friction between dust and gas in the upper disk
layers, due to electrostatic interactions between charged grains and the
ionized plasma.  Another intriguing possibility is that even the larger
interstellar grains are already porous aggregates
\citep{1989ApJ...341..808M,2002EAS.....4...31S} --- an effect which would
increase their surface area and therefore their drag coupling to the gas.
And finally, it could be possible that, for some unknown reason,
some disks are more turbulent than others.  The more turbulent disks then
remain flaring while the quiescent disks will become self-shadowed in a very
short time. If this is true, then this would imply that group II sources are
not evolutionarily linked to group I sources. However, since many
indications point toward such an evolutionary link, and since there seems to
be no good reason why some disks are more turbulent than others (other
properties remaining the same), we consider this scenario less likely.

\section{Conclusion}
We find that dust settling can change the shape of the photosphere of
protoplanetary disks on time scales of less than $10^3$ years at small radii
($\lesssim 10$AU) and $10^5$ years at larger radii. On a time scale shorter
than $10^6$ years a equilibrium state sets in in which settling is balanced
by turbulence.  In the equilibrium state the photosphere of the disk may
have a partly self-shadowed shape, so that the outer regions of the disk are
not anymore directly exposed to the stellar radiation. They are only
indirectly heated \revised{by infrared emission from other parts of the disk.}

Close to the star the settling in the upper layers of the disk proceeds much
faster than at large radii, and an equilibrium solution is therefore reached
more quickly. This means that the reduction in the photospheric height of
the disk ($\Hs(R)$) first happens at small radii, and then happens at
progressively larger radii, until after about $10^5$ to $10^6$ years the
equilibrium state has been reached globally. The transition from flaring
disk to half-flaring-half-self-shadowed disk therefore happens on the latter
time scale.

The flaring disks (before settling) are expected to have a reasonably strong
mid- to far-infrared excess, while the partly self-shadowed disks have a
much weaker mid- to far-infrared excess. Moreover, the partly self-shadowed
disks should become virtually undetectable in resolved scattered light
images beyond the self-shadowing radius, \revised{whereas non-self-shadowed
disks (early stages) should be much brighter. But even without
self-shadowing our calculations show that settling can cause dimming of the
disk in scattered light. This is by a `differentiation' process in which the
larger/heavier more reflective grains settle deeper into the disk, leaving
the smaller/lighter less reflective grains determining the optical
appearance of the surface of the disk.}

\begin{acknowledgements}
CPD wishes to thank C.~Grady for stimulating discussions that made us aware
of the observational evidence for shadowing in disks. We also thank
L.~Waters, Th.~Henning and J.~Cuzzi for very helpful discussions and advice.
\revised{Finally, we wish to thank the referee, R.~Lachaume, for his
extensive comments which have improved the manuscript considerably.}
\end{acknowledgements}


\begin{thebibliography}{35}
\expandafter\ifx\csname natexlab\endcsname\relax\def\natexlab#1{#1}\fi

\bibitem[{{Augereau} {et~al.}(2001){Augereau}, {Lagrange}, {Mouillet}, \& {M{\'
  e}nard}}]{augereau:2001}
{Augereau}, J.~C., {Lagrange}, A.~M., {Mouillet}, D., \& {M{\' e}nard}, F.
  2001, \aap, 365, 78

\bibitem[{{Bouwman} {et~al.}(2001){Bouwman}, {Meeus}, {de Koter}, {Hony},
  {Dominik}, \& {Waters}}]{bouwmanmeeus:2001}
{Bouwman}, J., {Meeus}, G., {de Koter}, A., {Hony}, S., {Dominik}, C., \&
  {Waters}, L.~B.~F.~M. 2001, \aap, 375, 950

\bibitem[{{Calvet} {et~al.}(1991){Calvet}, {Patino}, {Magris}, \&
  {D'Alessio}}]{calvetpatino:1991}
{Calvet}, N., {Patino}, A., {Magris}, G.~C., \& {D'Alessio}, P. 1991, \apj,
  380, 617

\bibitem[{{Chiang} \& {Goldreich}(1997)}]{chianggold:1997}
{Chiang}, E.~I. \& {Goldreich}, P. 1997, \apj, 490, 368+

\bibitem[{{Chiang} {et~al.}(2001){Chiang}, {Joung}, {Creech-Eakman}, {Qi},
  {Kessler}, {Blake}, \& {van Dishoeck}}]{chiangjoung:2001}
{Chiang}, E.~I., {Joung}, M.~K., {Creech-Eakman}, M.~J., {Qi}, C., {Kessler},
  J.~E., {Blake}, G.~A., \& {van Dishoeck}, E.~F. 2001, \apj, 547, 1077

\bibitem[{{Cuzzi} {et~al.}(2003){Cuzzi}, {Davis}, \&
  {Dobrovolskis}}]{2003Icar..166..385C}
{Cuzzi}, J.~N., {Davis}, S.~S., \& {Dobrovolskis}, A.~R. 2003, Icarus, 166, 385

\bibitem[{{Cuzzi} {et~al.}(1993){Cuzzi}, {Dobrovolskis}, \&
  {Champney}}]{cuzzidobrchamp:1993}
{Cuzzi}, J.~N., {Dobrovolskis}, A.~R., \& {Champney}, J.~M. 1993, Icarus, 106,
  102+

\bibitem[{{D'Alessio} {et~al.}(2001){D'Alessio}, {Calvet}, \&
  {Hartmann}}]{dalessiocalvet:2001}
{D'Alessio}, P., {Calvet}, N., \& {Hartmann}, L. 2001, \apj, 553, 321

\bibitem[{{Dubrulle} {et~al.}(1995){Dubrulle}, {Morfill}, \&
  {Sterzik}}]{dubmorster:1995}
{Dubrulle}, B., {Morfill}, G., \& {Sterzik}, M. 1995, Icarus, 114, 237

\bibitem[{{Dullemond}(2002)}]{dullemond:2002}
{Dullemond}, C.~P. 2002, \aap, 395, 853

\bibitem[{{Dullemond} {et~al.}(2001){Dullemond}, {Dominik}, \&
  {Natta}}]{duldomnat:2001}
{Dullemond}, C.~P., {Dominik}, C., \& {Natta}, A. 2001, \apj, 560, 957

\bibitem[{{Dullemond} {et~al.}(2003){Dullemond}, {van den Ancker}, {Acke}, \&
  {van Boekel}}]{dulancackboe:2003}
{Dullemond}, C.~P., {van den Ancker}, M.~E., {Acke}, B., \& {van Boekel}, R.
  2003, \apjl, 594, L47

\bibitem[{{Dullemond} {et~al.}(2002){Dullemond}, {van Zadelhoff}, \&
  {Natta}}]{dulvzadnat:2002}
{Dullemond}, C.~P., {van Zadelhoff}, G.~J., \& {Natta}, A. 2002, \aap, 389, 464

\bibitem[{{Eisner} {et~al.}(2003){Eisner}, {Lane}, {Akeson}, {Hillenbrand}, \&
  {Sargent}}]{eislanake:2003}
{Eisner}, J.~A., {Lane}, B.~F., {Akeson}, R.~L., {Hillenbrand}, L.~A., \&
  {Sargent}, A.~I. 2003, \apj, 588, 360

\bibitem[{{Grady} {et~al.}(1999){Grady}, {Woodgate}, {Bruhweiler}, {Boggess},
  {Plait}, {Lindler}, {Clampin}, \& {Kalas}}]{gradywoodbruh:1999}
{Grady}, C.~A., {Woodgate}, B., {Bruhweiler}, F.~C., {Boggess}, A., {Plait},
  P., {Lindler}, D.~J., {Clampin}, M., \& {Kalas}, P. 1999, \apjl, 523, L151

\bibitem[{{Honda} {et~al.}(2003){Honda}, {Kataza}, {Okamoto}, {Miyata},
  {Yamashita}, {Sako}, {Takubo}, \& {Onaka}}]{hondakataza:2003}
{Honda}, M., {Kataza}, H., {Okamoto}, Y.~K., {Miyata}, T., {Yamashita}, T.,
  {Sako}, S., {Takubo}, S., \& {Onaka}, T. 2003, \apjl, 585, L59

\bibitem[{{Kenyon} \& {Hartmann}(1987)}]{kenyonhart:1987}
{Kenyon}, S.~J. \& {Hartmann}, L. 1987, \apj, 323, 714

\bibitem[{{Kim} {et~al.}(1993){Kim}, {Martin}, \&
  {Hendry}}]{1993AAS...182.0813K}
{Kim}, S., {Martin}, P.~G., \& {Hendry}, P. 1993, Bulletin of the American
  Astronomical Society, 25, 806

\bibitem[{{Mannings} \& {Sargent}(1997)}]{mannsarg:1997}
{Mannings}, V. \& {Sargent}, A.~I. 1997, \apj, 490, 792+

\bibitem[{{Mathis} {et~al.}(1977){Mathis}, {Rumpl}, \& {Nordsieck}}]{mrn:1977}
{Mathis}, J.~S., {Rumpl}, W., \& {Nordsieck}, K.~H. 1977, \apj, 217, 425

\bibitem[{{Mathis} \& {Whiffen}(1989)}]{1989ApJ...341..808M}
{Mathis}, J.~S. \& {Whiffen}, G. 1989, \apj, 341, 808

\bibitem[{{Meeus} {et~al.}(2003){Meeus}, {Sterzik}, {Bouwman}, \&
  {Natta}}]{meeussterz:2003}
{Meeus}, G., {Sterzik}, M., {Bouwman}, J., \& {Natta}, A. 2003, \aap, 409, L25

\bibitem[{{Meeus} {et~al.}(2001){Meeus}, {Waters}, {Bouwman}, {van den Ancker},
  {Waelkens}, \& {Malfait}}]{meeuswatersbouw:2001}
{Meeus}, G., {Waters}, L. B. F.~M., {Bouwman}, J., {van den Ancker}, M.~E.,
  {Waelkens}, C., \& {Malfait}, K. 2001, \aap, 365, 476

\bibitem[{{Millan-Gabet} {et~al.}(2001){Millan-Gabet}, {Schloerb}, \&
  {Traub}}]{millanschl:2001}
{Millan-Gabet}, R., {Schloerb}, F.~P., \& {Traub}, W.~A. 2001, \apj, 546, 358

\bibitem[{{Miyake} \& {Nakagawa}(1995)}]{miyakenaka:1995}
{Miyake}, K. \& {Nakagawa}, Y. 1995, \apj, 441, 361

\bibitem[{{Natta} {et~al.}(2001){Natta}, {Prusti}, {Neri}, {Wooden}, \&
  {Grinin}}]{nattaprusti:2001}
{Natta}, A., {Prusti}, T., {Neri}, R., {Wooden}, D., \& {Grinin}, V.~P. 2001,
  \aap, 371, 186

\bibitem[{{Przygodda} {et~al.}(2003){Przygodda}, {van Boekel}, {{\` A}brah{\`
  a}m}, {Melnikov}, {Waters}, \& {Leinert}}]{przygoddaboek:2003}
{Przygodda}, F., {van Boekel}, R., {{\` A}brah{\` a}m}, P., {Melnikov}, S.~Y.,
  {Waters}, L.~B.~F.~M., \& {Leinert}, C. 2003, \aap, 412, L43

\bibitem[{Safronov(1969)}]{safronov-book}
Safronov, V.~S. 1969, Evolution of the Protoplanetary Cloud and Formation of
  the Earth and Planets (Moskow: Nauka Press (in Russian)), (English
  translation: NASA TTF-677, 1972)

\bibitem[{{Stepnik} {et~al.}(2002){Stepnik}, {Jones}, {Abergel}, {Bernard},
  {Boulanger}, \& {Ristorcelli}}]{2002EAS.....4...31S}
{Stepnik}, B., {Jones}, A.~P., {Abergel}, A., {Bernard}, J.~P., {Boulanger},
  F., \& {Ristorcelli}, I. 2002, EAS Publications Series, Volume 4, Proceedings
  of Infrared and Submillimeter Space Astronomy", held 11-13 June, 2001.~
  Edited by M.~Giard, J.P.~Bernanrd, A.~Klotz, and I.~Ristorcelli.~ EDP
  Sciences, 2002, pp.31-31, 4, 31

\bibitem[{{Takeuchi} \& {Lin}(2002)}]{takeuchilin:2002}
{Takeuchi}, T. \& {Lin}, D.~N.~C. 2002, \apj, 581, 1344

\bibitem[{{Takeuchi} \& {Lin}(2003)}]{takeuchilin:2003}
---. 2003, \apj, 593, 524

\bibitem[{{van Boekel} {et~al.}(2003){van Boekel}, {Waters}, {Dominik},
  {Bouwman}, {de Koter}, {Dullemond}, \& {Paresce}}]{vanboekelwaters:2003}
{van Boekel}, R., {Waters}, L.~B.~F.~M., {Dominik}, C., {Bouwman}, J., {de
  Koter}, A., {Dullemond}, C.~P., \& {Paresce}, F. 2003, \aap, 400, L21

\bibitem[{{Voelk} {et~al.}(1980){Voelk}, {Morfill}, {Roeser}, \&
  {Jones}}]{voelkmorroejon:1980}
{Voelk}, H.~J., {Morfill}, G.~E., {Roeser}, S., \& {Jones}, F.~C. 1980, \aap,
  85, 316

\bibitem[{{Weidenschilling}(1977)}]{weidenschilling:1977}
{Weidenschilling}, S.~J. 1977, \mnras, 180, 57

\bibitem[{{Weidenschilling}(1997)}]{weid1997}
---. 1997, Icarus, 127, 290

\end{thebibliography}
\end{document}